\documentclass[10pt]{elsarticle}
\usepackage{graphicx}
\usepackage{amsfonts, amsmath, amssymb}
\usepackage{bm}
\usepackage{cancel}
\usepackage{color}
\usepackage{enumerate}
\usepackage{float}
\usepackage{hyperref}
\usepackage[english]{babel}
\usepackage[margin=1in]{geometry}
\usepackage{mathtools}
\usepackage{setspace}
\usepackage{amsthm}
\usepackage{subfigure}
\usepackage{lineno}
\usepackage{array,multirow}
\newcolumntype{K}[1]{>{\centering\arraybackslash}p{#1}}
\usepackage{comment}

\makeatletter
\newcommand*{\rom}[1]{\expandafter\@slowromancap\romannumeral #1@}
\makeatother

\makeatletter
\renewcommand*\env@matrix[1][\arraystretch]{%
	\edef\arraystretch{#1}%
	\hskip -\arraycolsep
	\let\@ifnextchar\new@ifnextchar
	\array{*\c@MaxMatrixCols c}}
\makeatother
\graphicspath{{../Figs/final/}}
\newcommand{\V}[1]{\bm{\mathrm{#1}}}

\def \grad{\nabla}

\date{}

\makeatletter
\def\ps@pprintTitle{%
	\let\@oddhead\@empty
	\let\@evenhead\@empty
	\def\@oddfoot{}%
	\let\@evenfoot\@oddfoot}
\makeatother

\begin{document}
	
	\begin{frontmatter}

\title{Quantifying the COVID19 infection risk due to droplet/aerosol inhalation }

\author[1,2]{Rahul Bale\corref{mycorrespondingauthor}}
\ead{rahul.bale@riken.jp}
\author[3]{Akiyoshi Iida}
\author[4]{Masashi Yamakawa}
\author[1,2]{ChungGang Li}
\author[1,2]{Makoto Tsubokura}
\address[1]{RIKEN Center for Computational Science, Kobe 6500047, Japan}
\address[2]{Department of Computational Science, Kobe University, Kobe,  Japan. }
\address[3]{Department of Mechanical Engineering, Toyohashi Institute of Technology, Toyohashi, Japan }
\address[4]{Department of Mechanical Engineering, Kyoto Institute of Technology, Kyoto, Japan }
\cortext[mycorrespondingauthor]{Corresponding author}


\begin{abstract}
The dose-response model has been widely used for quantifying the risk of infection of airborne diseases like COVID-19. The model has been used in the room-average analysis of infection risk and analysis using passive scalars as a proxy for aerosol transport. However, it has not been employed for risk estimation in numerical simulations of droplet dispersion. In this work, we develop a framework for the evaluation of the probability of infection in droplet dispersion simulations using the dose-response model. We introduce a version of the model that can incorporate the higher transmissibility of variant strains of SARS-CoV2 and the effect of vaccination in evaluating the probability of infection. Numerical simulations of droplet dispersion during speech are carried out to investigate the infection risk over space and time using the model.   {The advantage of droplet dispersion simulations for risk evaluation is demonstrated through the analysis of the effect of ambient wind and humidity on infection risk.}
\end{abstract}
\end{frontmatter}	
	
\section{Introduction}	
Dispersal of infectious pathogens such as viruses through airborne sputum droplets can lead to the rapid transmission of diseases in the general population posing a great risk to public health. There are several minor and major infectious diseases that are transmitted through virus-carrying sputum droplets. A benign example of airborne disease is the common cold, while more severe examples include H1N1 influenza, severe acute respiratory syndrome (SARS), middle east respiratory syndrome (MERS), and coronavirus disease 2019 (COVID-19)\cite{wei16,gral11}. Since its outbreak in 2019, COVID-19 has transformed into the most destructive pandemic in over a century. The evidence on COVID-19 so far suggests that the possible modes of transmission of SARS-CoV-2 include respiratory droplets and aerosols, direct person-to-person contact and contact with surfaces (fomite mode of transmission). Direct person-to-person contact transmission and fomite transmission can be controlled by appropriate hygiene practices. However, controlling airborne transmission is far more challenging. Therefore, airborne transmission is likely the primary reason for turning COVID-19 into a global pandemic \cite{morawska2020,asadi2020}. The viruses carrying sputum droplets are generated not only during violent expiratory events like coughing and sneezing, but they are also generated during routine respiratory activities like speaking, singing, and breathing. This, coupled with the fact that  SARS-CoV2 is transmitted during presymptomatic and asymptomatic phases of COVID-19\cite{ferretti2020}, increases the infectiousness and the rapid spread of SARS-CoV2.  

As part of developing effective strategies to mitigate the transmission of COVID-19 or any airborne disease in the future, there is a need for quantification of the risk of infection under various social conditions. The well-mixed room approximation is one of the approaches that has been used by researchers for the estimation of infection of risk. By treating the concentration of carbon dioxide in exhaled air as a proxy for virion concentration in the ambient environment, the Wells-Riley model has been used for risk estimation in an indoor environment based on room-scale averages\cite{rudnick2003,issarow2015}.   {Given that the mechanisms of removal of aerosols and carbon dioxide in indoor ventilation systems are very different, this assumption is reasonable only in limited circumstances.} Recent works on this quantification of COVID-19 have also used a similar approach and applied it to indoor spaces such as bathrooms, laboratories and offices wherein the viral shedding of occupants was taken as the source of virions in room-average analyses \cite{smith20,augenbraun20,kolinski21}. These approaches provide a measure of the average infection risk for the whole environment under consideration rather than the local risk of infection for each occupant.   {In a study of the role of ventilation on the transmission of airborne respiratory pathogens, Li et al. \cite{li2022} adopted a mechanistic model for evaluating short and long-range airborne pathogen exposures in which it was shown that improved ventilation lowers exposure to airborne pathogens. The mechanistic model adopted by Li et al. provides useful information like the effect of room-averaged ventilation, effects airflow direction, mixing, temperature, and humidity cannot be considered. } Yang et al. developed a framework for risk estimation between individuals during casual conversation\cite{yang20}. The method involves the estimation of viral concentration as a function of space and time by using passive scalars as a proxy for aerosols ejected during speech. The risk of infection is then estimated by evaluating the viral particles that are likely to be within the breathing zone of a susceptible individual. A similar method has been adopted by Singhal et al. for risk estimation during face-to-face conversations\cite{singhal2021}.  These methods assume that the aerosols are well-mixed with exhaled air close to the source of aerosol ejection. However, the well-mixed aerosol assumption is reasonable for high aerosol concentration but not when the aerosol concentration is low. Moreover, the approach cannot account for the possible local variations in aerosol concentration due to changes in environmental factors like temperature and humidity.  The aerosolization of small and medium-sized droplets is influenced by the humidity and temperature of the ambient environment.  Direct measurement of aerosols and droplets that are likely to be within the breathing zone of a susceptible individual can provide a  better estimate of infection risk. The focus of the present work is the estimation of infection risk by direct measurement of droplets and aerosols in the breathing zone of susceptible individuals using numerical simulation of droplet dispersion. To that end, we extend the dose-response model for droplet dispersion simulations for expiratory events like coughing, sneezing, speaking, etc. Using numerical simulations of droplet dispersion, infection risk during face-to-face conversation is investigated in this work.   {Furthermore, we also investigate the effect of wind and humidity on the infection probability during casual conversations.}

\section{Estimation of Infection Probability}
	The risk of infection due to inhalation of viruses is commonly quantified using the dose-response model\cite{wells1955,watanabe2010,watanabe2012}. The model assumes that the number of viral particles needed, on average, to infect an individual is $ N_0 $. Assuming that the infection is a Poisson process, the probability of infection can be written as
\begin{equation}
	P = 1 - e^{(-\frac{N}{ N_0})},
	\label{eqn:Prisk}
\end{equation}
where $ N $ is the total number of virions inhaled. In order to compute the probability of infection $ P $, both $  N $ and $ N_0 $ have to be estimated. A range of values have been reported in the literature for $ N_0 $. Prentiss et al. have estimated a range of 322-2012 based on superspreading events at the early stage of COVID-19 pandemic \cite{prentiss20},  other similar studies provide estimates of the value that vary between 100 and 1000 \cite{kolinski21,augenbraun20}. In this study, we choose a value of 900 which lies within the range of values reported in the literature. 

The number of virions inhaled depends on the total duration of exposure, $ T $, the amount of air inhaled by a person breathing at the rate of B, and the local concentration of virions,   {$ C(x,t) $}, in the breathing zone of the person. Therefore, the relationship of  $ N $, with $ T, B \;\& \;C $ can be expressed as  
  {
	\begin{equation}
		N(x,T) = B \int_{0}^{T}C(x,t)dt.
		\label{eqn:N-v1}
\end{equation}}
While breathing rate is relatively constant for a given physical activity, the local concentration is transient and can vary with time. Depending on the type of physical activity, the breathing rate can vary from moderate values ($ 0.45 $ m$ ^3 $/hr) to values as high as $ 3.3 $ m$ ^3 $/hr \cite{binazzi06,united1989,shimer1995}. The breathing rate for sedentary activities like quiet breathing, speaking at normal voice and singing are in a similar range: $ 0.54\pm0.21 $ m$ ^3 $/hr,  $ 0.54\pm0.21 $ m$ ^3 $/hr and  $ 0.61\pm0.4 $ m$ ^3 $/hr, respectively\cite{binazzi06}. For moderate physical activities like cycling and climbing stairs the breathing rate varies between  $ 1.3-1.5 $ m$ ^3 $/hr, and for heaving activities like climbing with load, cross country skiing, the range is $ 2.5-3.3 $ m$ ^3 $/hr \cite{united1989}. For the present study, we choose a value of $ B=0.5 $ m$ ^3 $/hr which assumes that the subject at risk of infection is not involved in any strenuous activity.  With the value of $ B $ known, the estimation of $ N(x,t) $ depends on the estimation of $ C(x,t) $. There have been two main approaches of estimating $ C(x,t) $ in literature. The first approach is based on a well-mixed room averaged approximation\cite{prentiss20,kolinski21}. In this approach, the virions emitted by an infected person are assumed to be well-mixed due to air circulation and mixing within the domain under consideration. The change in virion concentration in the room is tracked over time, but the concentration is assumed to be uniform throughout the domain under investigation. An alternate approach proposed by Yang et al.\cite{yang20} involves estimation of $ C(x,t) $ over time and space. In this method, assuming that the aerosol-laden air is well-mixed at the point of ejection during expiratory events such as speaking, singing, coughing, etc., a passive scalar is used to model the transport of aerosols in direct numerical simulation or large eddy simulations.  Taking passive scalar concentration as a proxy for virion concentration, an expression of $ C(x,t) $ is obtained. 

\begin{figure}[ht]
	\centering
	\includegraphics[width=0.8\textwidth]{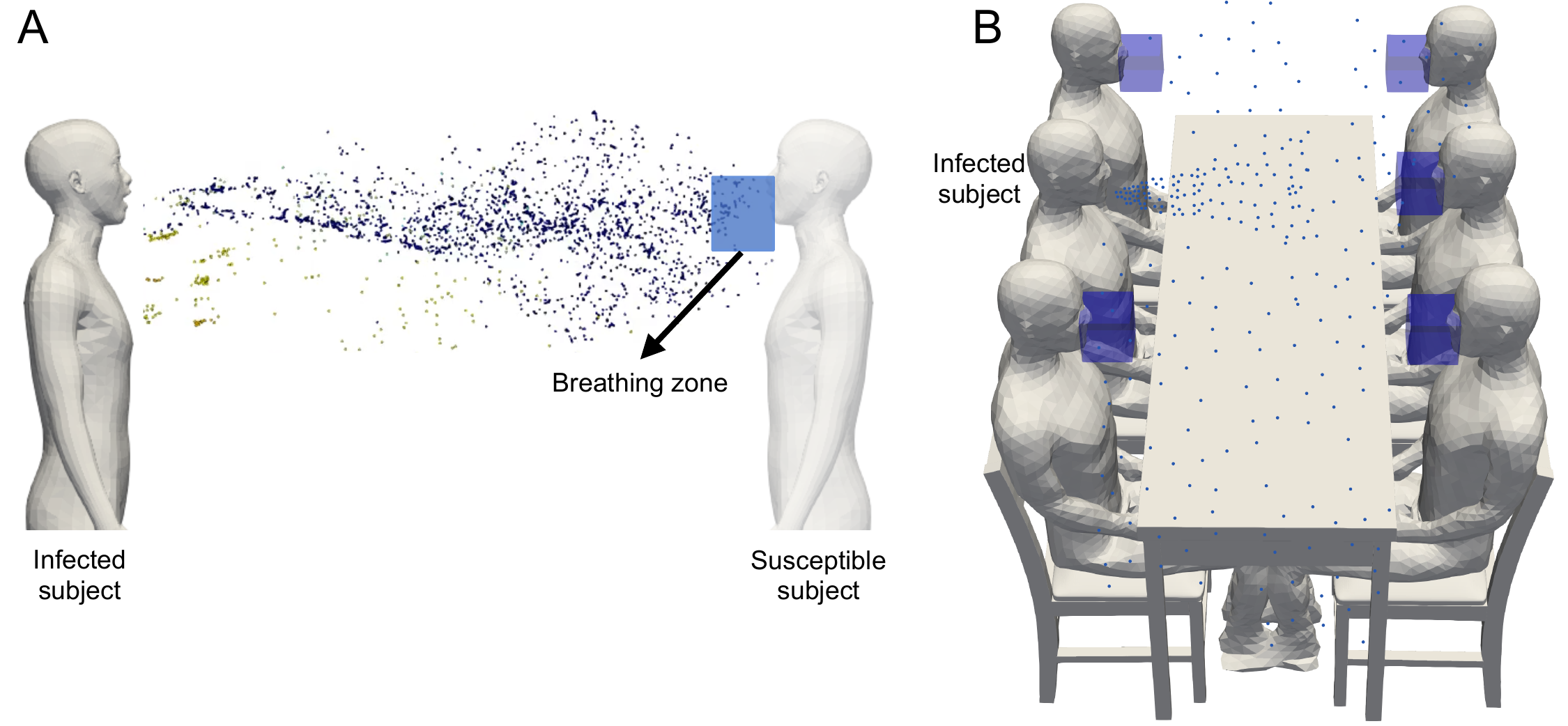}
	\caption{A schematic of droplet dispersion (A) during face to face conversation between two subjects, (B) between a group of people seated at a table. The region highlighted by a blue box around the mouth and nose of a susceptible subject, termed the breathing zone, is used to track droplets that are likely to be inhaled.}
	\label{fig:breathing-zone-schematic}
\end{figure}

In contrast to the two methods available in the literature, in this work, we develop a new model to estimate virion inhalation as a function of space and time in droplet dispersion simulations.  As opposed to the well-mixed approximation of aerosol-laden air in the passive scalar approach, in simulations modeling direct droplet dispersion, the dynamics of aerosol/droplet transport and evaporation are considered. In contrast to aerosol transport, the dynamics of transport and mixing of droplets can be different due to the larger inertia of droplets. Aerosolization of small and medium droplets, depending on the ambient temperature and humidity, can significantly alter local aerosol and virion concentration. Therefore, the results of droplet dispersion simulations can be qualitatively different from those of the passive scalar approach.  

For the evaluation of infection probability in droplet dispersion simulations, we estimate $ N(x,t)  $ by tracking the number of droplets in a spatial region defined such that the air within this region is likely to be inhaled by a susceptible subject.  Henceforth, we shall refer to this region as the breathing zone. A schematic of the breathing zone is shown in Fig.~\ref{fig:breathing-zone-schematic}. For simplicity, we have chosen a small rectangular region around the mouth and nose to represent the breathing zone. The shape of the region could in principle be arbitrary. An example of how the breathing zone is defined in two scenarios is shown in Fig.~\ref{fig:breathing-zone-schematic}. The first scenario is a face-to-face conversion between two persons. Second is a conversion between a group of people seated at a table. In all such situations, one or more infected subjects are modeled as the source of droplets/aerosols while the remaining subjects are treated as being susceptible to infection. A separate breathing zone is defined for each subject who might be susceptible to inhaling the virus-laden droplets and getting infected. A separate infection probability of each subject would be evaluated by tracking the droplets in the respective breathing zones. We choose a $ 10\times10\times15cm $ sized rectangular region to model the breathing zone for the present study.   {This is approximately of the order of exhalation and inhalation volume of the speaking model considered in this work.} The longer side of the box is along the nose-ground direction. With regards to the placement of the breathing zone, the center of one of the edges of the top face of the box is placed at the bottom edge of the nose and an adjacent face is placed in contact with the outermost surface of the mouth of a subject. Lastly, the breathing zone placement is such that its entire volume is outside the subject's interior.  

The  number of virions the subject is likely to inhale depends on the local concentration of virions in the breathing zone.  The local concentration can be written as  
\begin{equation}
	C(x,t) =  \frac{n(x,t)}{\upsilon_{\text{B}}}, 
\end{equation}
where $ n(x,t) $ is the instantaneous count of virions at the time instant t in the breathing zone and $ \upsilon_{\text{B}} $ is the volume of the breathing zone. The instantaneous virion count $ n(x,t) $ is the product of the total ejection volume of droplets and aerosols, $ \upsilon_{d}^{0}(x,t) $,  in the breathing zone and the viral load or viral density $ \lambda_{v} $ (copies/m$ ^3 $). The superscript $ ^0 $ in $ \upsilon_{d}^{0}(x,t) $ is used to imply that the volume of the droplets in question is the volume at the time of ejection from the mouth of the infected subject. Even as a droplet evaporates, its instantaneous volume, $ \upsilon_{d} $, decreases but its injection volume remains unchanged. Therefore, the total virion count of a given droplet remain constant as a droplet evaporates and aerosolizes. The  average viral load of SARS-CoV2 in the sputum is $ 7\times 10^6 $ copies/ml and the maximum reported value is $ 2\times 10^9 $ copies/ml\cite{wolfel20}. For the analysis in the present work we choose the value $ 7\times 10^6 $ copies/ml.

The local virion concentration can now be expressed as a function of viral load and droplet volume 
\begin{equation}
	C(x,t) = \frac{\upsilon_{d}^{0}(x,t)\lambda_{v}}{\upsilon_{\text{B}}}. 
\end{equation}
Substituting the above expression into Eq.~\ref{eqn:N-v1} we obtain the following expression for $ N $
\begin{equation}
	N(x,T) =  \frac{B\lambda_{v}}{\upsilon_{\text{B}}} \int_{0}^{T} \upsilon_{d}^{0}(x,t)dt.	
	\label{eqn:N-v3}
\end{equation}
In activities like speaking and singing over prolonged periods of time, the rate of droplet generation and dispersion are expected to reach a quasi-steady-state. The injection droplet volume that enters the breathing zone would reach a steady state value $ \overline{\upsilon}_{d}^{0} $. Therefore, under steady-state  conditions the above equation can be simplified to 
\begin{equation}
	N(x,T) =  \frac{B\lambda_{v} \overline{\upsilon}_{d}^{0}(x) T}{\upsilon_{\text{B}}} .	
	\label{eqn:N-steady-state}
\end{equation}
This equation is applicable to quasi-steady processes like singing and speaking and it is not applicable to transient situations like sneezing and coughing. We must resort to Eq.~\ref{eqn:N-v3} for transient cases. \\

  {In order to contrast the present approach with the approach based on the  well-mixed approximation, wherein passive scalars are used as a proxy for aerosols, we can arrive at the expression for $ N $ adopted in the work of Yang et al. \cite{yang20} from Eq.~\ref{eqn:N-v1}.  The local virion concentration based on the local droplet volume fraction $ \phi(x,t) $ (droplet volume/air volume) can written as $ C(x,t) = \phi(x,t) \lambda_{v} $. As the passive scalar is used a proxy for the 'well-mixed' droplet aerosols, the droplet volume fraction is a function of the scalar concentration $ Y_s(x,t) $ and the volume fraction at source $ \phi^o $. Therefore, the local virion concentration is given by $ C(x,t) =  (Y_s(x,t)/Y_{s}^{o}) \phi^o \lambda_{v} $ , where $ Y_{s}^{o} $ is the scalar concentration at source (typically $ Y_{s}^{o}=1 $). Substituting this into Eq.~\ref{eqn:N-v1},  expression for $ N $ for unsteady and steady state processes can be obtained 
	
	\begin{align}
		N(x,T) = &  \frac{ B \phi^o \lambda_{v}}{Y_{s}^{o}} \int_{0}^{T} Y_s(x,t)dt,  \label{eqn-N_phi_a}\\
		N(x,T) = & \frac{ B \phi^o \lambda_{v} \overline{Y}_s(x)T}{Y_{s}^{o}},	
		\label{eqn-N_phi_b}
	\end{align}
	where $ \overline{Y}_{s}(x) $ is the time averaged scalar concentration. The volume fraction of droplets ejected from the mouth during conversations has been reported to range between  $ 2\times 10^{-9} $ and $ 1 \times 10^{-8} $ \cite{leung2020}. In the work of Yang et al. \cite{yang20} and  Singhal et al. \cite{singhal2021}, a value of $ 6\times 10^{-9} $ has been used for casual conversations, and  $ 1\times 10^{-8} $ for loud conversation by  Yang et al. }

\noindent\textbf{Risk after vaccination and due to variant strains}: Most of the major vaccines for COVID-19 have reported high efficacy in preventing infection and very high efficacy in preventing severe disease and hospitalization. For example, vaccines by Pfizer, Moderna and AstraZeneca  have been reported to have vaccine efficacy of $ 95\% $\cite{polack2020}, $ 94.1\% $\cite{baden2021} and $ 81.5\% $\cite{emary2021}, respectively. From the viewpoint of evaluating the infection probability of a vaccinated subject, the effect of a vaccine may be interpreted as an increase in the minimum number of virions needed to infect a person. This assumption implies that for exposure to small doses of virions, the infection probability will be very low or negligible. However, even an infected person, if exposed to very high doses of virions, may be at risk of infection. 

As SARS-CoV2 has spread through populations, it has mutated into several variants\cite{cdc-variants}. It has been reported that the transmissibility of some of the variants of SARS-CoV2 is higher than that of the original strain \cite{campbell2021,davies2021}. For example, the B.1.1.7 strain (alpha variant)  has been reported to be $ 29\% $ more transmissible than the original strain\cite{campbell2021} and the B.1.617.2 strain (delta variant) has been estimated to be 43 to 90$ \% $ more transmissible than B.1.1.7 strain\cite{davies2021}. Within the framework of the infection risk model considered in this work, the higher transmissibility of a given variant could be interpreted as due to two reasons. First, the viral load of the variant strains could be higher. Second, the minimum virion dose needed for infection $ N_0 $ for a variant could be lower. The higher transmissibility could be due to one or a combination of these two factors. Under this assumption, the effect of vaccines and variant strains on the probability of infection can be incorporated into the risk model given by Eq.~\ref{eqn:Prisk} as follows.

\begin{equation}
	P = 1 - e^{(-\alpha\frac{N}{ N_0})},
	\label{eqn:Prisk-vaccine}
\end{equation}
where $ \alpha $ is a factor that accounts for higher transmissibility of variant strains, lower risk of infection for vaccinated individuals, and with $ \alpha = 1 $ the above equation falls back to the original form in Eq.~\ref{eqn:Prisk}. For the case where a person is vaccinated with a vaccine of efficacy $ \eta_{vc} $,  it can be shown that $ \alpha = 1-\eta_{vc} $. The details of derivation of  $ \alpha $  can be found in Section.~\ref{asec:vacc}.  The effect of vaccination may be interpreted as an increase in the minimum number of virions $ N_0 $ needed to infect a vaccinated person. If the efficacy of a vaccine is $ 100\% $ then the $ P=0 $. But, when vaccine efficacy is 0, Eq.~\ref{eqn:Prisk-vaccine} returns to the original form of $ P $ in Eq.~\ref{eqn:Prisk}.  For a variant strain with a higher transmissibility factor $ \tau $, we can write $ \alpha = \tau $. For the alpha, strain the transmissibility factor  is $ \tau = 1.29 $\cite{campbell2021}, and for the delta strain it is $ \tau = 2.45 $\cite{davies2021} (assuming 90$ \% $ higher transmissibility compared to alpha variant). For the original strain the transmissibility factor is $ \tau = 1 $, in which case Eq.\ref{eqn:Prisk-vaccine} return to the original form in Eq.\ref{eqn:Prisk}.

\section{Methods}
\subsection{Governing Equations}
The flow solver used in the present work for carrying out the numerical simulation of droplet dispersion is made of an Eulerian reference frame for solving the fluid flow and species transport equations and a Lagrangian frame for solving the droplet dynamics model. The equations of motion of mass, momentum, energy and species transport can be expressed in compact notation as 
\begin{equation}
	\frac{\partial\rm{U}}{\partial t} +\nabla\cdot \mathbf{F} = \mathbf{S}.
	\label{eq:ge}
\end{equation}
Here, $\rm{U}$, $\mathbf{F}$ and $\mathbf{S}$ represent the primitive flow variables, the combined convective and diffusive terms, and the source terms, respectively \cite{poinsot2005theoretical}. The primitive variables vector and the flux vector are expanded below. 
\begin{equation}
	\mathbf{U}=\left(\begin{array}{c}
		\rho \\
		\rho u_{1} \\
		\rho u_{2} \\
		\rho u_{3} \\
		\rho e \\
		\rho Y_{k}
	\end{array}\right), \quad F_{i}=\left(\begin{array}{c}
		\rho u_{i} \\
		\rho u_{i} u_{1}+P \delta_{i 1}-\mu A_{i 1} \\
		\rho u_{i} u_{2}+P \delta_{i 2}-\mu A_{i 2} \\
		\rho u_{i} u_{3}+P \delta_{i 3}-\mu A_{i 3} \\
		\rho(\rho e+P) u_{i}-\mu A_{i j} u_{j}+q_{i} \\
		\rho u_{i} Y_{k}-\rho \hat{u}_{i}^{k} Y_{k}
	\end{array}\right)
	\label{eqn-UF}
\end{equation}
where the density and viscosity are represented by $ \rho $ and $ \mu $, respectively.  $\mathbf{u}$, $e$ and $P$ are the velocity, the total specific energy and the pressure, respectively.  The components of the velocity along the principle directions $1,2,3$ are given by $ ( u_1, u_2, u_3 ) $. The vapor phase of water from the liquid sputum is modeled as passive scalar species along with O$ _2 $ and N$ _2 $. The mass fraction of the species indexed $ k $ is represented by $ Y_k $ and $\hat{u}^{k}_i$ is the corresponding diffusion velocity of the $ k^{th} $ species. $ 	\mathbf{q} = - \lambda \grad T, $ is the heat flux where  $T \& \lambda$ represent temperature and thermal diffusivity, respectively. The density and pressure are constrained together by the state equation $ P=\rho RT $, in which $ R $ is the gas constant and $ T $ is the temperature.  The total specific energy is given by  
\begin{equation}
	e = \frac{P}{\gamma - 1} + \frac{1}{2} u_i u_i,
\end{equation}
where $\gamma$ is the ratio of the gas specific heat capacities.  The diffusion velocity  $\hat{\V{u}}^k Y_k$ of the $k^{\textrm{th}}$ species is defined in terms of the species diffusivity $D_k$, the relationship is given by 
\begin{equation}
	\hat{\V{u}}^k Y_k = D_k \grad Y_k.
\end{equation}
  {The contribution to the source comes from the buoyancy term and the evaporation of the droplets. As the sputum droplet motion is driven by the flow generated in the mouth and not vice-versa, the contribution of the droplet motion to the momentum of the carrier fluid is not expected to be significant. Therefore, the same is neglected. The total mass of the sputum ejected due to droplets smaller than 100$ \mu m $ is $ \sim 10^{-8} kg $.  The sputum droplets at the point of ejection from the mouth are spread across a volume of a width of at least $2 cm^2 $. The mass of air in this region is $\sim 8\times 10^{-6} g. $   Even instant evaporation of all the droplets over such a volume would contribute only about $ \sim0.125\% $ mass to the carrier gas. Therefore, the contribution of the droplet ($ d_d< 100\mu m$) to carrier gas mass due to evaporation is negligible. As for droplets with $d_d >100 \mu m $, the time scale of evaporation relative to a time scale of velocity change due to gravity is very long. As a result, large droplets rapidly settle on the ground before they evaporate, at which point these droplets and their contribution to the system can be ignored. \\
	The evaporation of droplets is strongly influenced by the partial pressure at the surface of the droplet, which depends on the local mass fraction of the vapor phase of the droplet. Small changes in the local mass fraction due to evaporation of the droplet can affect the evaporation of the droplet, creating a feedback loop. Therefore, the contribution of the droplet evaporation is modeled through the source term to the species transport equation of the vapor phase of the droplet. } The source term vector is given by
\begin{equation}
	\mathbf{S} =  \begin{pmatrix} 
		{0} \\
		{(\rho-\rho_0)g_1} \\ 
		{(\rho-\rho_0)g_2} \\ 
		{(\rho-\rho_0)g_3} \\  
		{(\rho-\rho_0)g_i u_i} \\ 
		{S_{\rho Y_{k}} }
	\end{pmatrix}.
\end{equation}
$ \rho $ and $ \rho_0 $ are the local and far field ambient density, respectively and $ \mathbf{g} $ is the acceleration due to gravity (eg. $ \mathbf{g}=(0,0,-9.81)m/s^2 $).  Of the species source terms $ S_{\rho Y_{k}} $, the non-droplet vapor species are zero. 
\subsection{Droplet Model}
The widely used single droplet model is adopted in this work for modeling the sputum droplet dynamics. The droplets are modeled as discrete Lagrangian entities which are coupled with the Eulerian fluid flow equations for a weak-two-way coupling. The droplet transport and evaporation are influenced by the conditions of the ambient air, but the flow field is not affected by the droplets except the species of the vapor phase of the liquid droplet. The transport of the droplets is modeled by 

\begin{equation}
	\begin{aligned}
		\frac{d \mathbf{x}_d}{d t} &= \mathbf{u}_d, \\
		\frac{d \mathbf{u}_{d}}{d t} &=\frac{3 C_{D}}{4 d_{d}} \frac{\rho}{\rho_{d}}\left(\mathbf{u}-\mathbf{u}_{d}\right)\left|\mathbf{u}-\mathbf{u}_{d}\right|+\mathbf{g} ,
		\label{eqn-dropletU}
	\end{aligned}		
\end{equation}
where $ \mathbf{x}_d $ and $ \mathbf{u}_d $ are the position and velocity of an individual droplet, respectively. $ d_d $  and $ \rho_d $ are the droplet diameter and liquid density of the droplet, respectively.    {$  C_D $ is the drag coefficient expressed a function of the droplet Reynolds number $ Re_d = \rho \textrm{max}(|\mathbf{u} - \mathbf{u}_d|) d_d/\mu$, and is given by
	\begin{equation}\label{eqn-cd}
		C_d = \begin{cases}
			\frac{24}{Re_d}\left( 1+1/6 Re_{d}^{2/3} \right) & Re_d < 1000, \\
			0.424 & Re_d > 1000,      
		\end{cases}  
	\end{equation}
	Other forces such as the Staffman lift force, buoyancy force, virtual mass force, Basset force, Magnus force, etc., can influence the droplet velocity. However, for particles of density in the order 1000 $kg/m^3 $, size in the range of 1$ \mu m$ to 100 $ \mu m $ suspended in a gas, the magnitude of these additional forces relative to the drag force is negligible \cite{zheng2020, mahian2019}. }

The droplet evaporation, influenced by the ambient air's velocity, humidity and temperature, is 
\begin{equation}
	\begin{aligned}
		\frac{d T_{d}}{d t} &=\frac{N u}{3 P r} \frac{c_{p}}{c_{l}} \frac{f_{1}}{\tau_{d}}\left(T-T_{d}\right) + \frac{1}{m_{d}}\left(\frac{d m_{d}}{d t}\right) \frac{L_{V}}{c_{p, d}} \\
		\frac{d m_{d}}{d t} &=-\frac{m_{d}}{\tau_{d}}\left(\frac{S h}{3 S c}\right) \ln \left(1+B_{M}\right)
	\end{aligned}
\end{equation}
The temperature $ T_d $ is updated by tracking the convective heat transfer with the ambient air and the evaporative heat loss.  { The mass rate of change of the droplets is influenced by the local vapor fraction and air velocity which are encapsulated in mass transfer number $ B_m =(Y_{v,s} - Y_v)/(1-Y_{v,s})$ and the Sherwood number $ Sh = 2 + 0.552 Re_{d}^{1/2}Sc^{1/3} $. 
	Here, $ Y_{v,s} $ and $ Y_{v} $ are the vapor mass fraction of the droplet at the surface and far field, respectively, and  $ m_d $ and $ L_V $ are the mass of the droplet and the latent heat of evaporation at the droplet temperature, respectively.} $ c_p $ and $ c_l $ are the specific heat at a constant pressure of the ambient air and the specific heat capacity of the liquid droplet, and $ \tau_d $ is the response time of the droplet. A unit lewis number assumption is applied wherein $ Pr=Sc $ and $ Pr $ is the Prandtl Number defined as the ratio of momentum and thermal diffusivity. Further details of the various terms involved in the droplet model can be found in the work of Bale et al.\cite{bale20b,bale2021}. The sputum droplet is assumed to be composed of water and virions. Evaporation of the stupum droplets stop when all the volatile component of the droplet evaporates resulting in droplet nuclei composed of virions.  The coronavirus has been reported to be 0.08 to 0.12 $ \mu $m in size \cite{masters2006}. Assuming that a droplet is composed of $ O(10^2) $ virions, the droplet nuclei diameter is limited to  1 $ \mu $m independent of a droplets initial diameter (nucleation of the droplet to smaller diameter will have negligible influence on its transport characteristics). Therefore, the evaporation of a droplet is stopped once the droplet diameter reaches this limit.

\subsection{Droplet modeling parameters}
\begin{figure}[!t]
	\centering
	\subfigure[]{ 
		\includegraphics[width=0.325\textwidth]{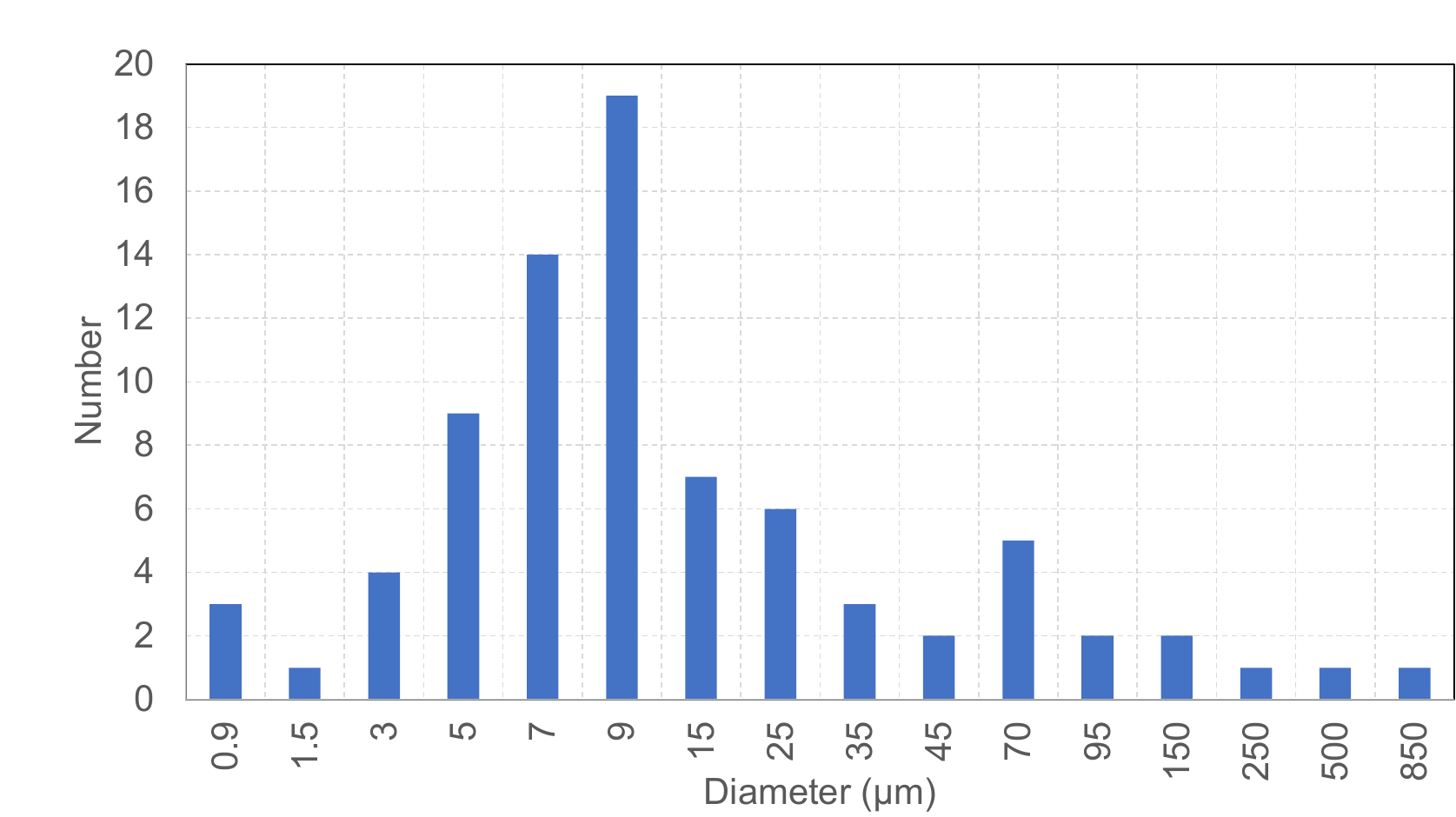} \label{fig:droplet-dist-soft}}	
	\subfigure[]{
		\includegraphics[width=0.325\textwidth]{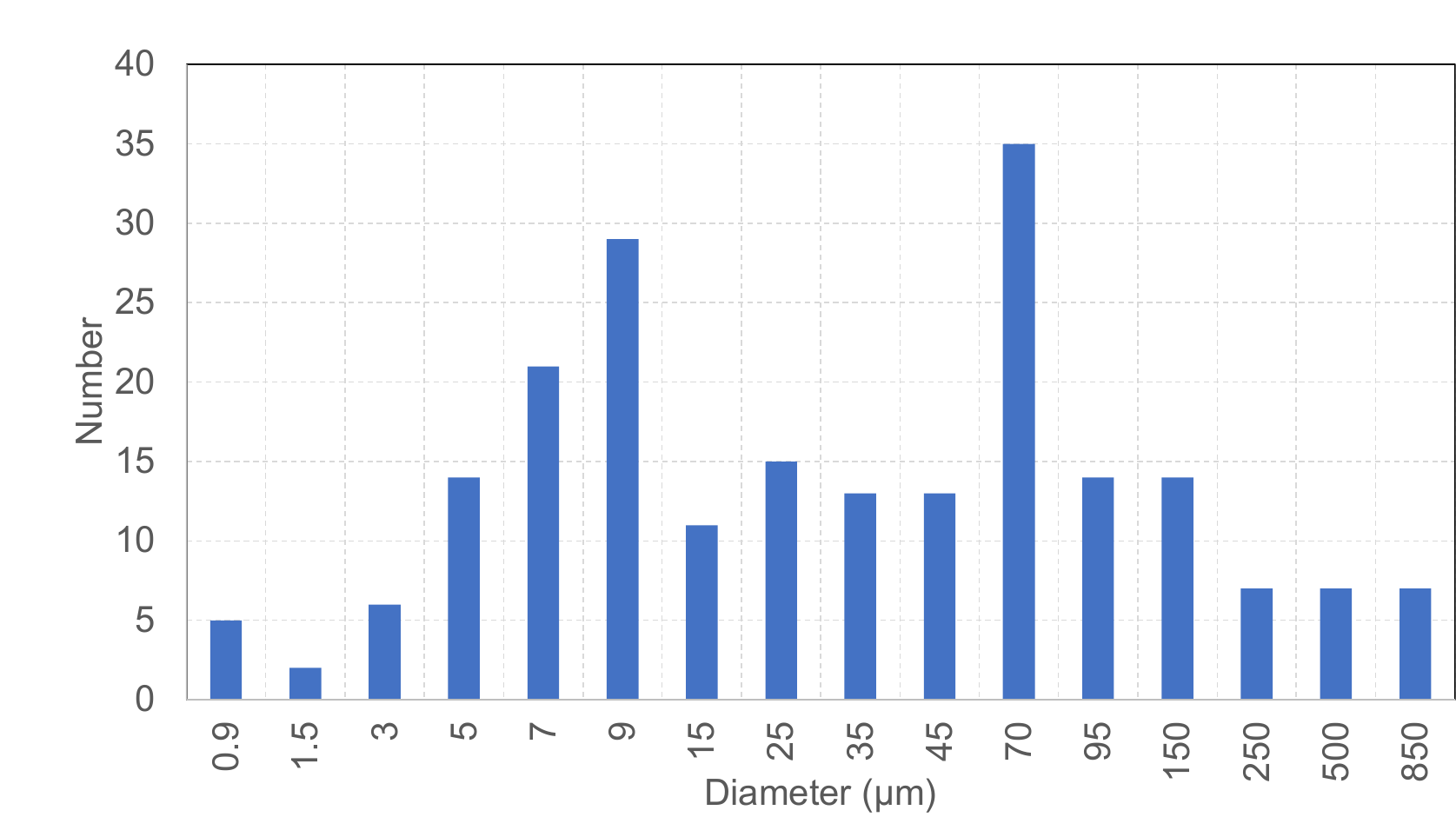}    }
	\subfigure[]{ 
		\includegraphics[width=0.28\textwidth]{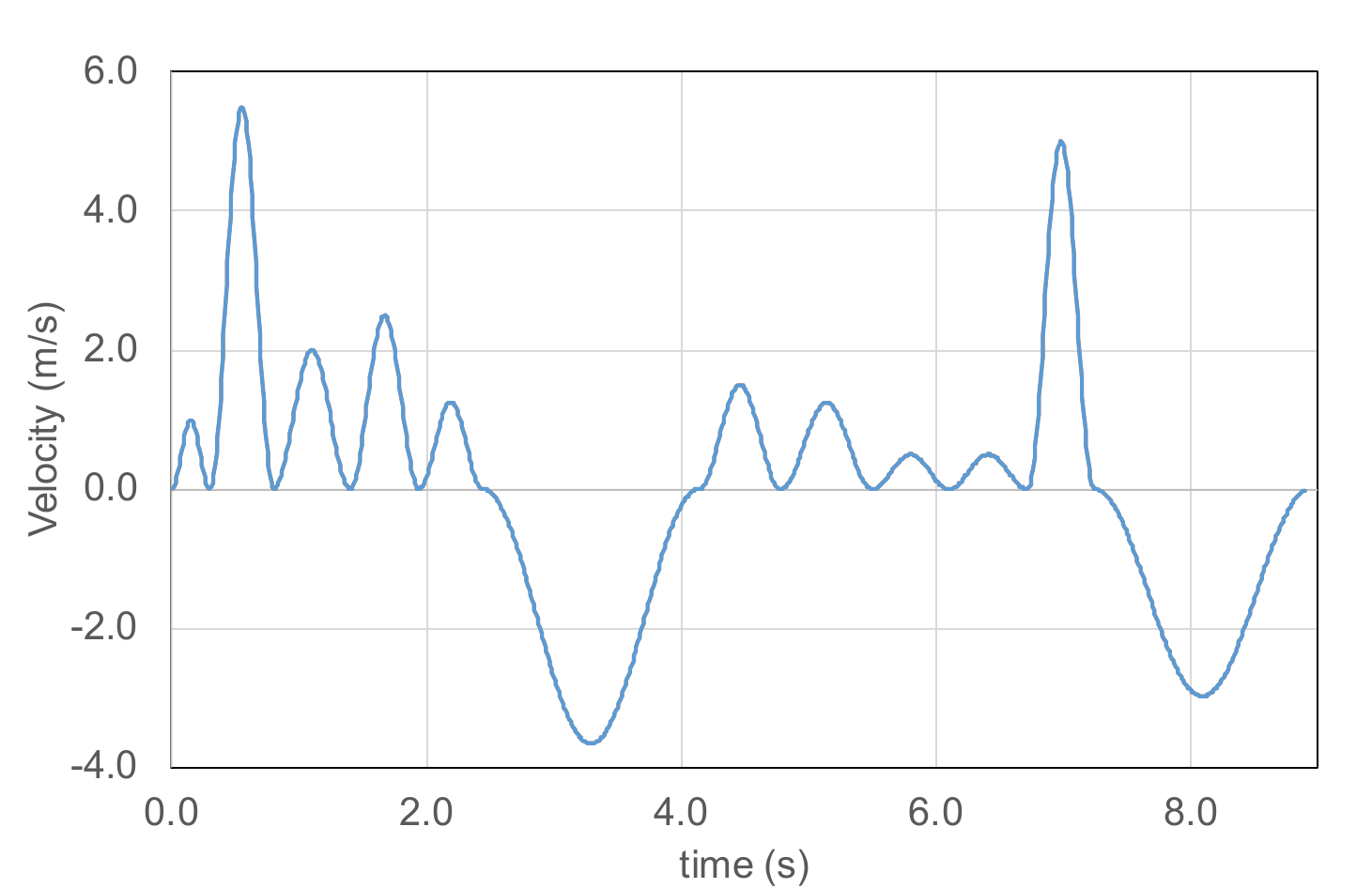}    }	
	\caption{ The distribution of droplet diameter used to model the droplet ejection while (a) speaking, and (b) loudly speaking. (c) The velocity profile of speaking flow over one cycle of speech which involves counting from 1 to 10 with two inhalation phases that balance of the volume of air exhaled.}
	\label{fig:droplet-dist-flow}
\end{figure}

Numerical simulation of droplet dispersion requires three main input parameters - a) the distribution of the droplet diameter and a count of the droplet number ejected from the mouth, b) the flow profile of the expiratory event in consideration such as speaking or coughing, c) the average area of the mouth opening. Data on the droplet size distribution and the number of droplets for speech as well as cough have been widely reported in literature\cite{loudon67,chao09,asadi2019,xie09}. The distribution and the droplet number reported for speaking in these studies varies significantly. The droplet concentration ($ \# $/L), a proxy for droplet number, for speaking reported by Duiguid, Loudon and Roberts, and Chao et al. are 3.72, 223.25 and 150.8\cite{duguid46,loudon67,chao09}, respectively. The difference in the reported droplet size distribution is also as disparate as the data on droplet number. The diameter corresponding to the highest droplet count reported by Loudon and Roberts was 6 $ \mu $m, this value in the study of Duguid and Chao et al. is about 12 $ \mu $m, on the other hand, Xie et al.\cite{xie09} report a value as high as 50 $ \mu $m.   {As there is so much variation in the number and distribution of droplet size in this work we adopt a combination of distribution Duguid\cite{duguid46} and Loudon $ \& $ Roberts \cite{loudon67}. The distribution adopted in this work essentially follows that of Duiguid \cite{duguid46} with a modification to the number count corresponding to $ 70\mu m $ diameter by doubling it to include the effect of the peak of Loudon $ \& $ Roberts distribution which occurs around $ 70 \mu m $. This distribution is shown in Fig.~\ref{fig:droplet-dist-soft}. Assuming that the droplet distribution reported for speaking in \cite{duguid46,loudon67} are for casual or soft speech, we alter the distribution in Fig~\ref{fig:droplet-dist-soft} for a loud speaking scenario as follows. The droplet count for diameters less than $ 20 \mu m $ are increased by a factor of 1.5, droplets larger than $ 70 \mu m $ are increased by a factor of 7, and the remaining droplets are increased by a factor that linearly increases from 1.5 to 7 as the diameter is increased from $ 20 \mu m $ to $ 70 \mu m $. The distribution of droplet diameter for loud speaking adopted for the numerical simulations in this work is shown in Fig.~\ref{fig:droplet-dist-flow}. }

For the flow generated from the mouth during speech, we adopt a sinusoidal model. The flowrate generated when counting from 1 to 10 is modeled as $ \dot{q}= A_i \sin^2(\pi t/T_i) $, where $ A_i $ is the amplitude and $ T_i $ is the period of the $ i^{th}$ utterance. After the word ’five’, the flow direction is reversed to model inhalation balancing the volume of air exhaled during speech counting from ’one’ to ’five’,. A similar inhalation is modeled after the word ’ten’. The velocity of the flow over an area of 6 cm$^2 $ over one cycle of counting from 1 to 10 including the two inhalations presented in Fig.~\ref{fig:droplet-dist-flow}. The period and amplitude of each utterance and the two inhalation phases of the speech flow can be deduced from the figure. The final parameter necessary to close the issue related to the boundary condition of droplet ejection is the area of the mouth opening. To the best of the authors’ knowledge, the information of the area of the mouth opening during speech is not available in the literature. An average mouth opening size of 4 cm$^2 $ was reported by Gupta et al.\cite{gupta09} for the situation of cough. Assuming that the mouth opening during speech is on average larger than the opening during cough, we choose a circular surface that is 6 cm$ ^2 $ in the area to model the mouth opening during speech. The droplets are injected at the circular mouth model into the domain at time instants that match the peak of the velocity of each utterance.

\subsection{Simulation setup}
\begin{figure}[!t]
	\centering
	\includegraphics[width=0.24\textwidth]{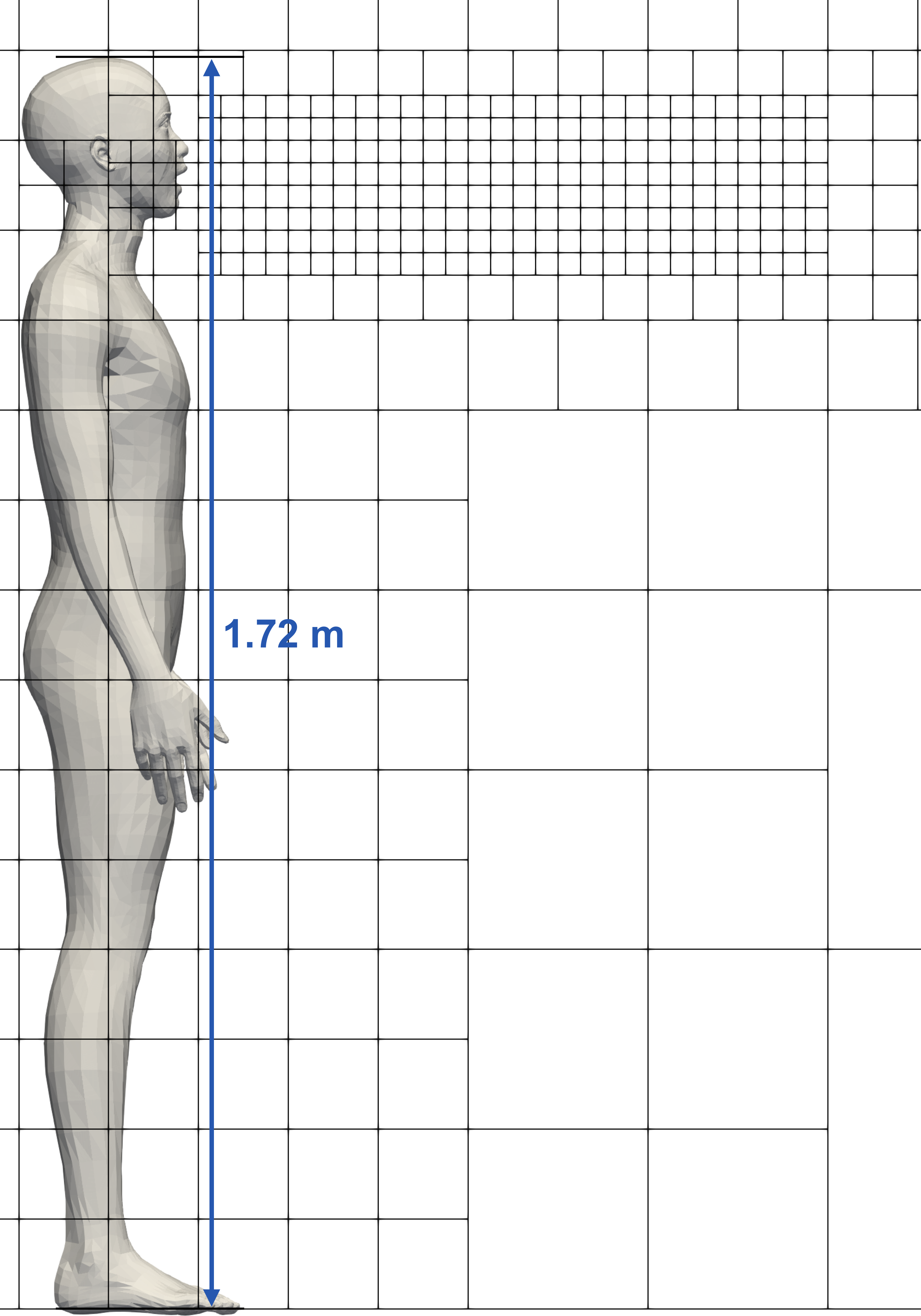}    
	\caption{Blocks of the numerical mesh employed in the present study. Subdivision of the blocks into $ 16 $ equal parts along each direction produces the cells.}
	\label{fig:mesh-geom}
\end{figure}

The simulation setup involves a human model standing in an upright position similar to the infected subject shown in Fig.\ref{fig:mesh-geom}. The numerical mesh employed in this work is also shown in the figure.   {A grid convergence study considering mesh spacing of 8 mm, 4 mm and 2 mm, up to a distance of 1m in front of the mouth, was carried out to determine the optimal mesh spacing for the present study. The results of the grid convergence study are presented in Appendix Fig~\ref{fig:mesh_study} The time-averaged centerline axial velocity and passive scalar concentration of the speech jet show a decaying trend with distance which does not significantly vary across the three mesh spacing considered. Therefore, we adopt the $ 2 $ mm mesh for the present study .} The numerical dissipation of the Roe-scheme\cite{roe81} used for the convective fluxes in our solver enable us to carry out implicit large eddy simulations (ILES)\cite{grinstein2007}. The human model is modeled with the immersed boundary condition\cite{li16} to impose no-slip and isothermal boundary conditions. The temperature of the human body surface is set to 300K to include the effects of buoyancy-driven flow by the human model, although the effect is not expected to be significant. The outer boundaries of the computational domain are treated with the slip boundary condition. The initial conditions for the simulation were set to the STP conditions. The temperature, pressure and relative humidity were set to 297 K, 101.3×103 Pa and 50$ \% $, respectively. The circular mouth geometry for modeling the speaking flow is placed 1 cm in front of the mouth of the human model. The flow generated by the speaking model is imposed on the circular mouth geometry. The relative humidity of the flow emanating from the mouth geometry is set to 90 $ \% $. The droplets generated from speaking are injected into the computational domain at random locations on the circular mouth geometry. The initial velocity of the droplets is set to 0, the droplets are to be driven by the flow from the time of injection into the domain. The initial temperature of the droplets is chosen to be 308 K matching the temperature of the interior of the human mouth and the density of the droplet is set to 1000 kg/s.   { The simulations are carried out for a duration of 75 s. After a period of 25 s from the beginning of the simulation, the total ejection volume of the droplets entering the breathing zone is time averaged for a period of 50 s to obtain $ \overline{\upsilon}^{0}_{d} $ }

\begin{figure}[!tb]
	\centering
	\includegraphics[width=0.5\textwidth]{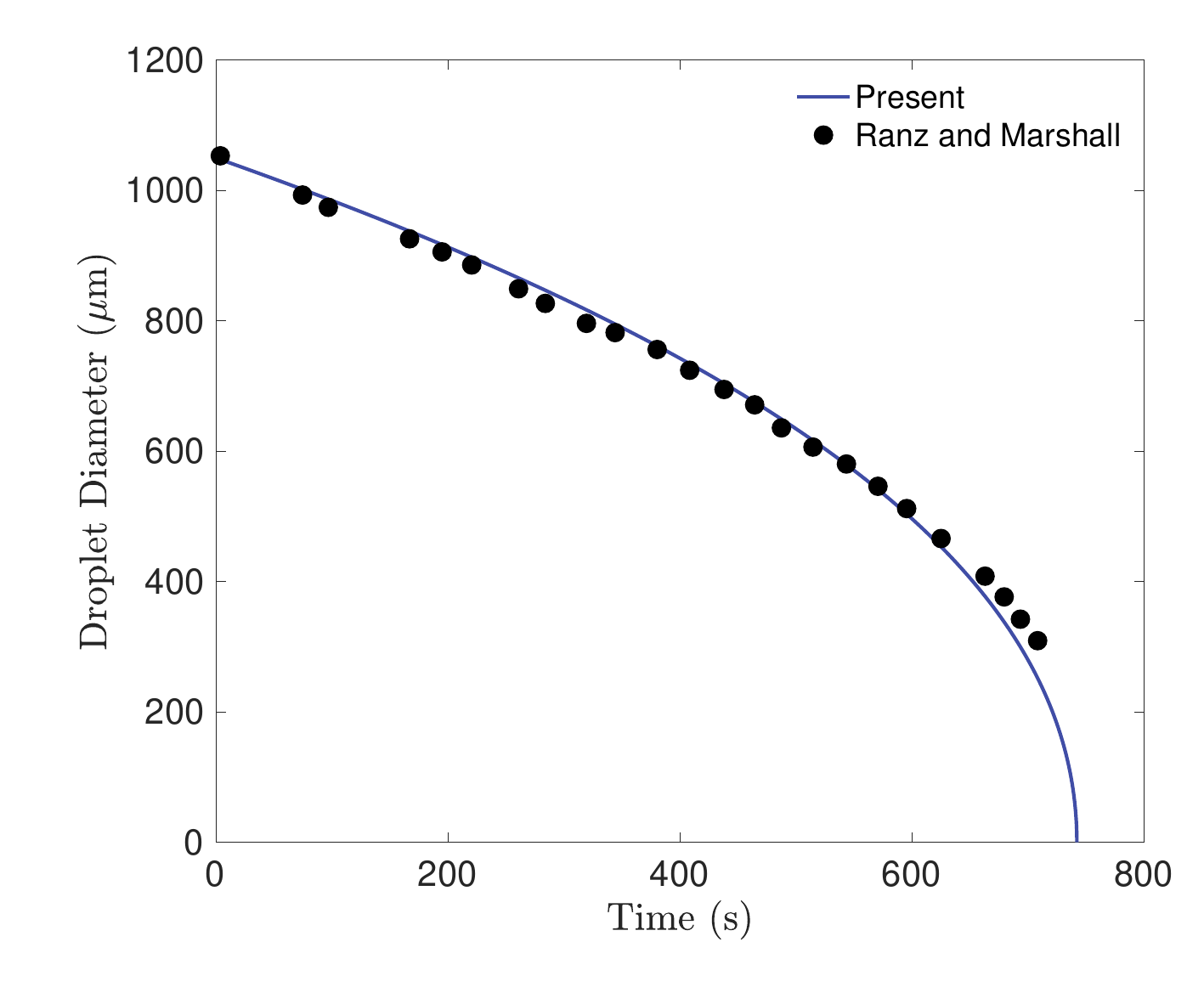}
	\caption{Comparison of droplet diameter as it evaporates with experimental data of Ranz and Marshall\cite{ranz52a,ranz52b}.}
	\label{fig:validation}
\end{figure}

\begin{figure}[!tb]
	\centering
	\subfigure[]{
		\includegraphics[width=0.48\textwidth]{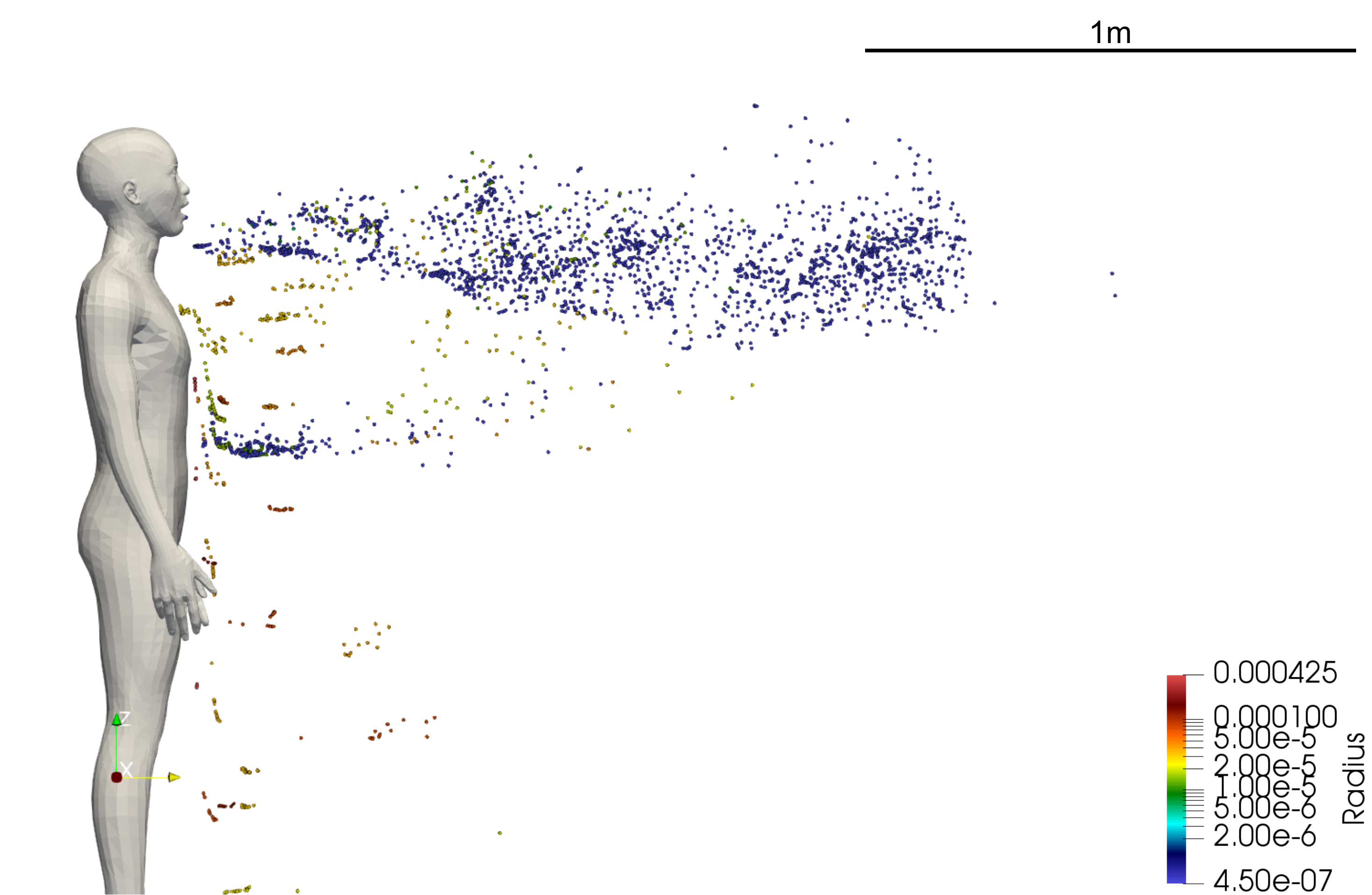}    }
	\subfigure[]{ 
		\includegraphics[width=0.48\textwidth]{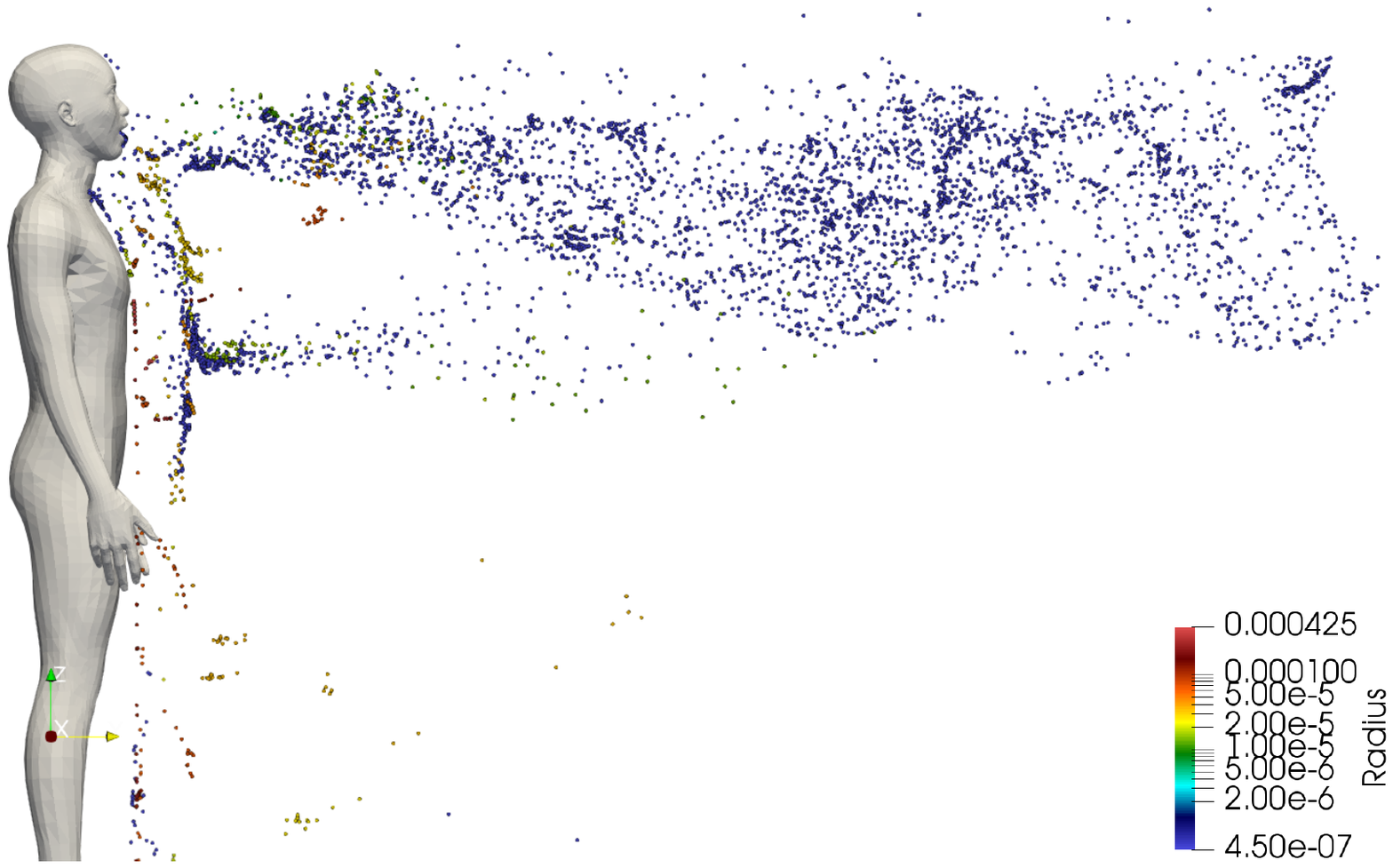}    }
	\caption{Dispersion of droplets during continuous speech at (a) $ t=12 $ s  (b) $ t = 24 $ s}
	\label{fig:droplet-viz}
\end{figure}

\section{Results}
\subsection{Validation}
A numerical simulation of the evaporation of a single isolated droplet was carried out to validate the droplet model. The experiment of Ranz and Marshall\cite{ranz52a,ranz52b}, in which the evaporation dynamics of a motionless droplet was carried, is used to validate our numerical simulation. The numerical setup mimicked the experimental setup wherein a motionless droplet of initial diameter $ d_d = 1050\;\mu $m is placed in an environment where the relative humidity and the temperature of the surrounding air were $ RH=0\% $ and $ T= 298$ K, respectively. The initial temperature of the droplet was $ T_d=282 $ K. After the exposure of the droplet to the surrounding environment, the evolution of the droplet's diameter as it evaporates is tracked and compared the experimental data of Ranz and Marshall. The comparison is plotted in Fig.~\ref{fig:validation} where it can be seen that there is excellent agreement between the simulation results and the experimental data.

\begin{figure}[!tb]
	\centering
	\subfigure[]{
		\includegraphics[width=0.48\textwidth]{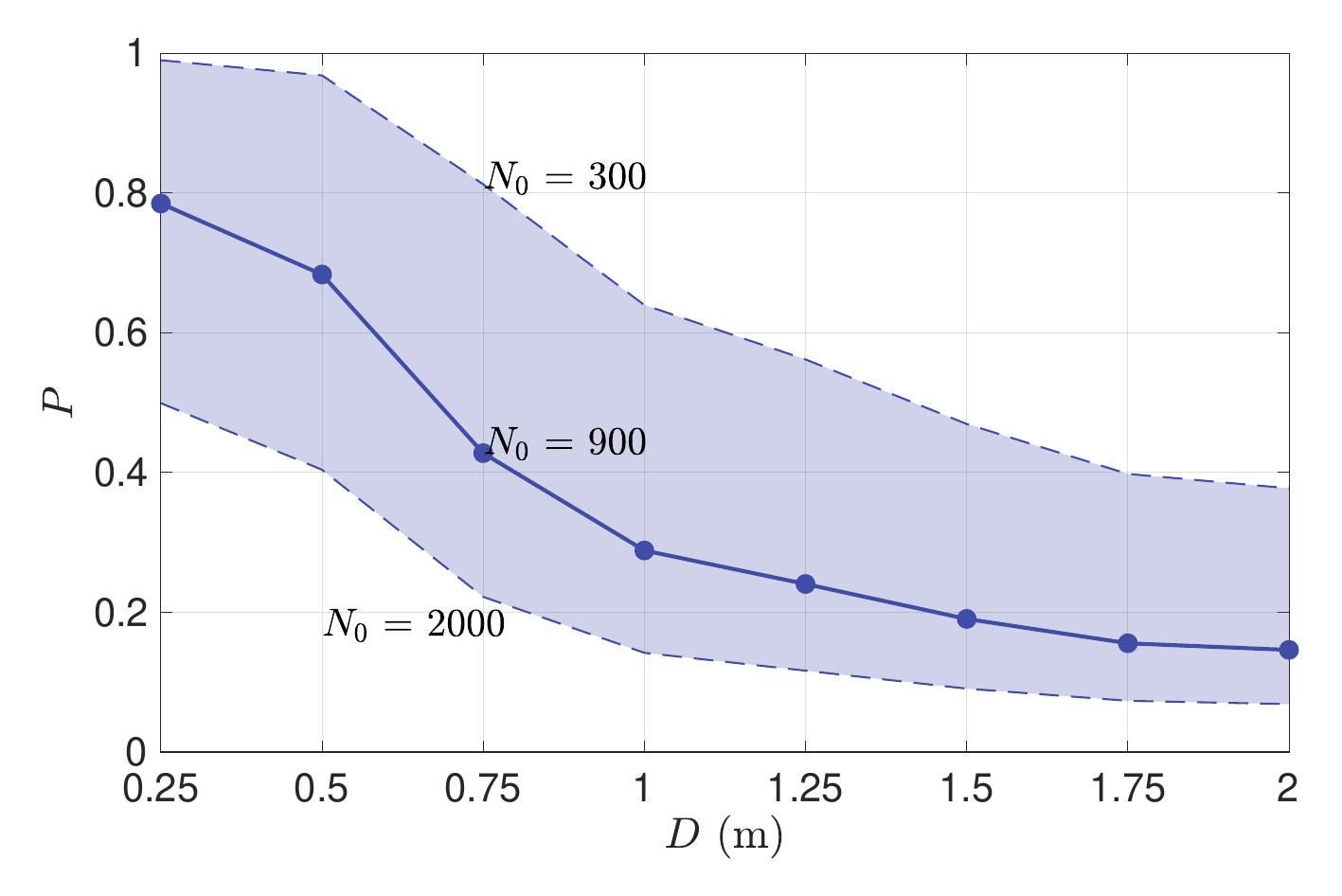}}   
	\subfigure[]{
		\includegraphics[width=0.48\textwidth]{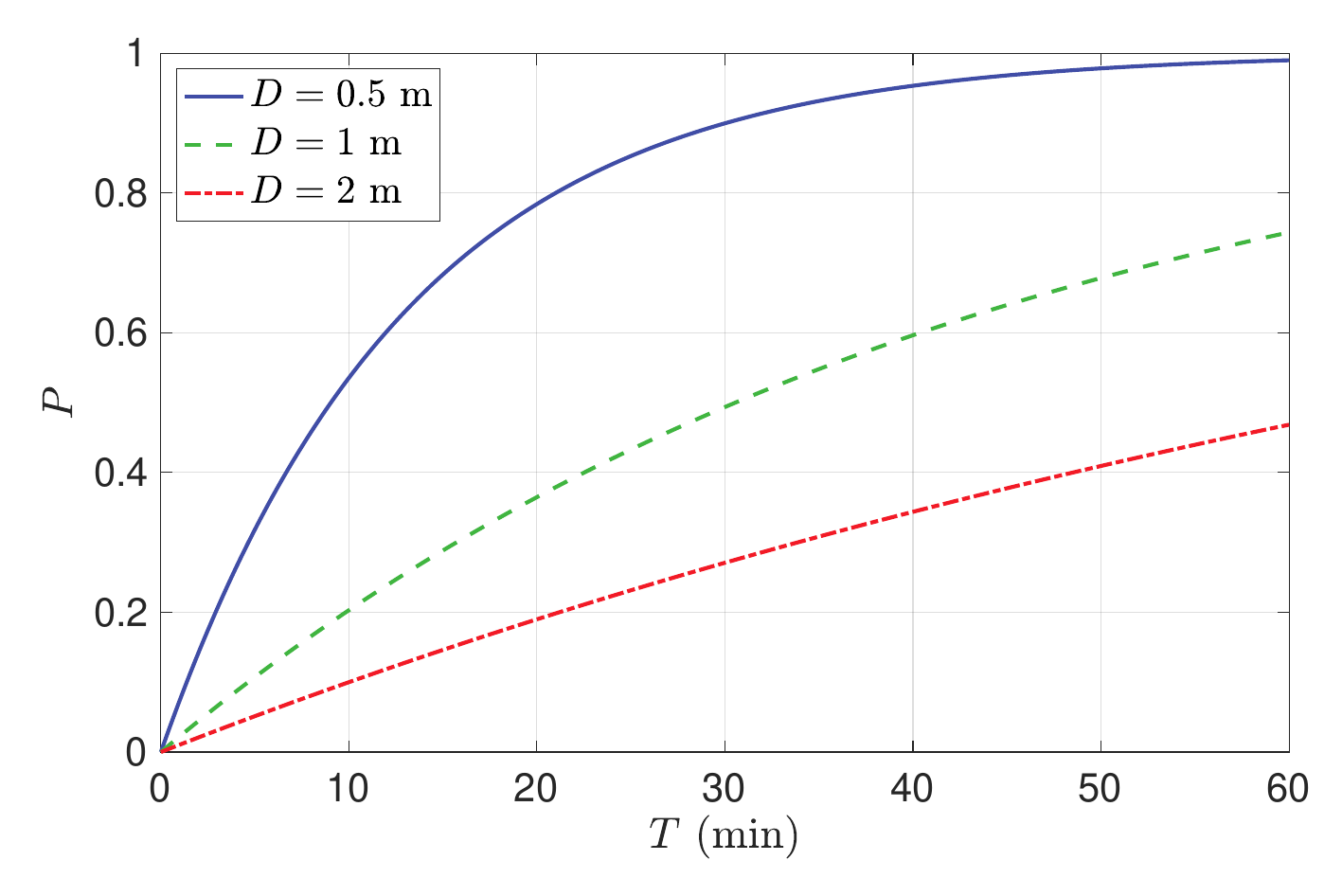}}
	\caption{(a)The probability of infection due to exposure for 15 mins plotted against distance during face-to-face conversation. (b) Evolution of $ P $ over time at distance of 0.5, 1.0, 2.0 m from the infection source.}
	\label{fig:RH50-PvsDist}
\end{figure}

\subsection{Infection risk during speech}
The infection risk model presented in this work is adopted to investigate the probability of infection of a susceptible subject who is in a face-to-face conversation with an infected individual. For this we carried out the numerical simulation of dispersion of droplets ejected by an infected subject in a standing pose assumed to be continuously speaking for the duration of the simulation. In the simulation, we model only the geometry of the infected subject, however, the geometry of the susceptible subject is not modeled.  The risk of infection of a virtual subject, whose dimensions are assumed to be identical to that of the infected subject, is evaluated at varying distances from the infected subject and also a function of time. The numerical simulation is carried out in a computational domain measuring $ 32\times 32\times 16 $ m$^3$ along x, y and z-axis, respectively. The human geometry is placed with the base of its feet at the center of the computational domain along x and y-axis and at the bottom along the z-axis. The details of the boundary and initial conditions of the setup can be found in the \textit{Methods} section. The details of the speaking model employed in this work are provided in the section \textit{Droplet modeling parameters}.  One cycle of the speaking model involves counting from 1 to 10 with two inhalation phases after the words '5' and '10', respectively, for mass balance (see Fig.~\ref{fig:droplet-dist-flow}). The speaking is modeled for the duration of the simulation by indefinitely looping the cycle of the speaking model. 

A visualization of instantaneous states of the dispersion of droplets at two time instants  is presented in Fig.~\ref{fig:droplet-viz}. The size of the droplets is indicated by the coloring scheme of the droplets. The largest droplets are colored red and the smallest blue. It is evident that many of the droplets larger than 20 $ \mu$m quickly settle on the ground under the influence of gravity. As the initial condition for droplet velocity is 0, gravity contribution dominates the velocity of the larger droplets. Therefore, the horizontal distance traversed by larger droplets is not significant when compared to smaller droplets. The influence of gravity on droplets smaller than 10$ \mu $ m is negligible because of aerosolization due to very short evaporation timescales and consequently the flow-induced drag forces dominate the small droplet and aerosol velocity. As a consequence, the smaller droplets and a small number of medium-sized droplets remain airborne and they are carried by flow generated by the speech. The dispersion of the aerosolized droplets in the horizontal direction over two instants of time is shown in Fig.~\ref{fig:droplet-viz}.

We next move on the investigation of infection probability of a virtual subject placed at different distances in front of the infected subject. The probability of infection at different distances in front of the infected subject for an exposure duration of $ T=15$ min is plotted in Fig.~\ref{fig:RH50-PvsDist}a. As for the infectious dose $ N_0 $, we have chosen a value of 900 which lies with the ranges values reported in the literature\cite{prentiss20,kolinski21,augenbraun20}. The variation of infection probability over distance exhibits a decaying profile. At distances less than 0.5m the probability of infection is greater than $ 70\% $, which rapidly decays to less than $ 20\% $ as the distance is increased to 2 m. It can be noted from Fig.~\ref{fig:droplet-viz} that the droplet concentration rapidly decays due to its dispersion in the vertical (z-axis) and in-plane direction (x-axis) as the droplets are advected away from the infection source thereby lowering the infection probability with distance.  The shaded region in the figure depicts the change in the infection risk if the infectious dose is changed from 300 to 2000. The infection risk at a given distance is lower for larger values of $ N_0 $ and vice-versa. It is interesting to note that the shaded region narrows as the distance from the infected person increases from 0.25 to 2 m. As the distance from the infection source increases, due to the dispersion of droplets, the virion concentration in the inhalation zone decreases which in turn decreases the number of inhaled virions. When the inhaled virion count is small enough, the magnitude of the infectious dose $ N_0 $ becomes less important resulting in the narrowing of the shaded region.   

The variation of infection probability overtime at distances $ D= (0.5, 1.0, 2.0) $ m is plotted in Fig.~\ref{fig:RH50-PvsDist}b. For large virion concentrations at closer distances like $ D=0.5 $m, the $ P $ rapidly increases and saturates to the maximum value (1). On the other hand, at a further distance, due to lower virion concentration, the rate of change of $ P $ is more gradual. At any given instant of time, say 10, 20, 30 min, etc., the plot provides the relative risk of maintaining different distances from a likely infected person. Focusing on the horizontal grid line corresponding to $ P=0.2 $., it can be seen that the exposure time required for 20$ \% $ probability of infection is approximately 3, 10 and 21 min for distances of 0.5, 1 and 2 m, respectively. 

\begin{figure}[!t]
	\centering
	\subfigure[]{
		\includegraphics[width=0.48\textwidth]{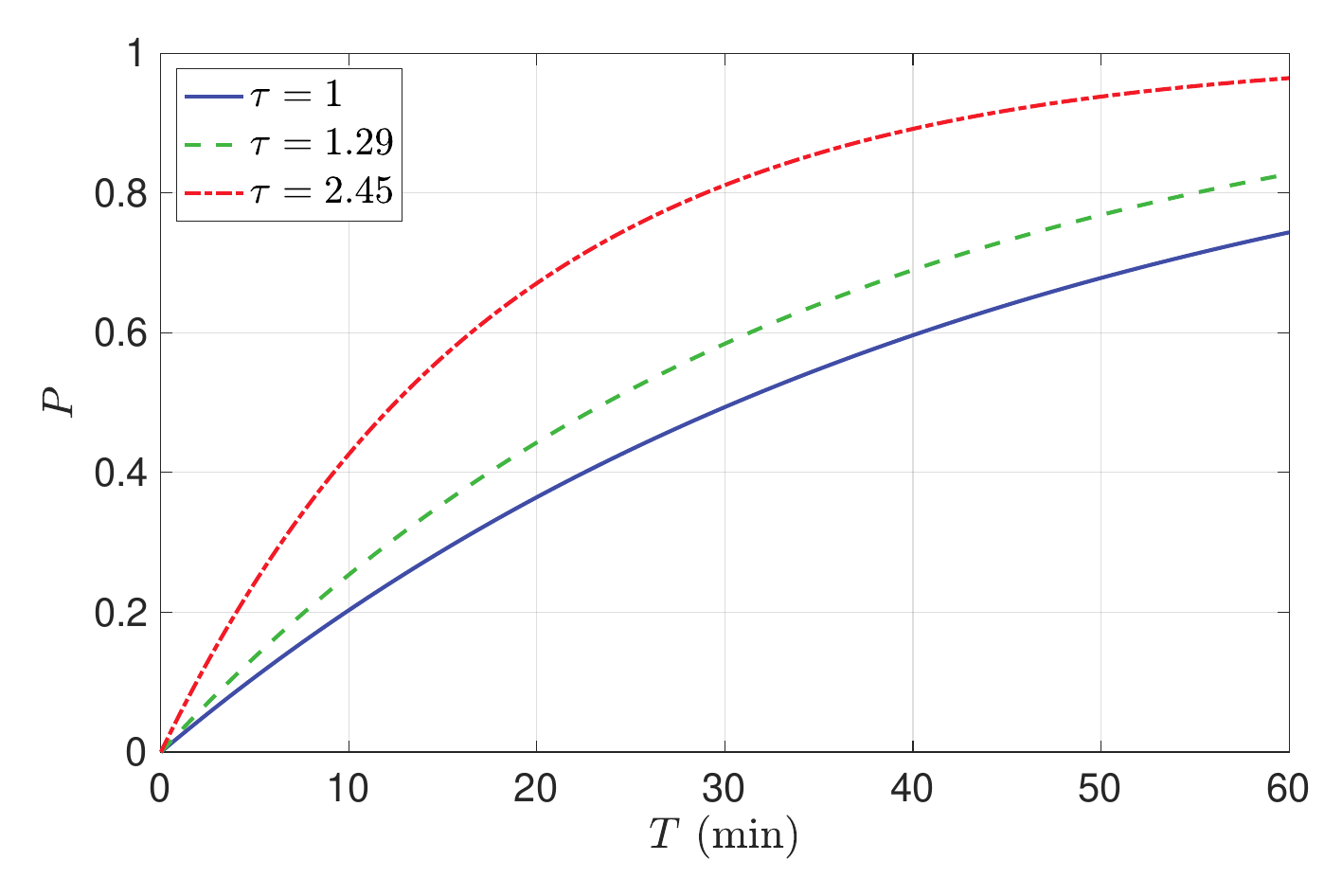}}   
	\subfigure[]{
		\includegraphics[width=0.48\textwidth]{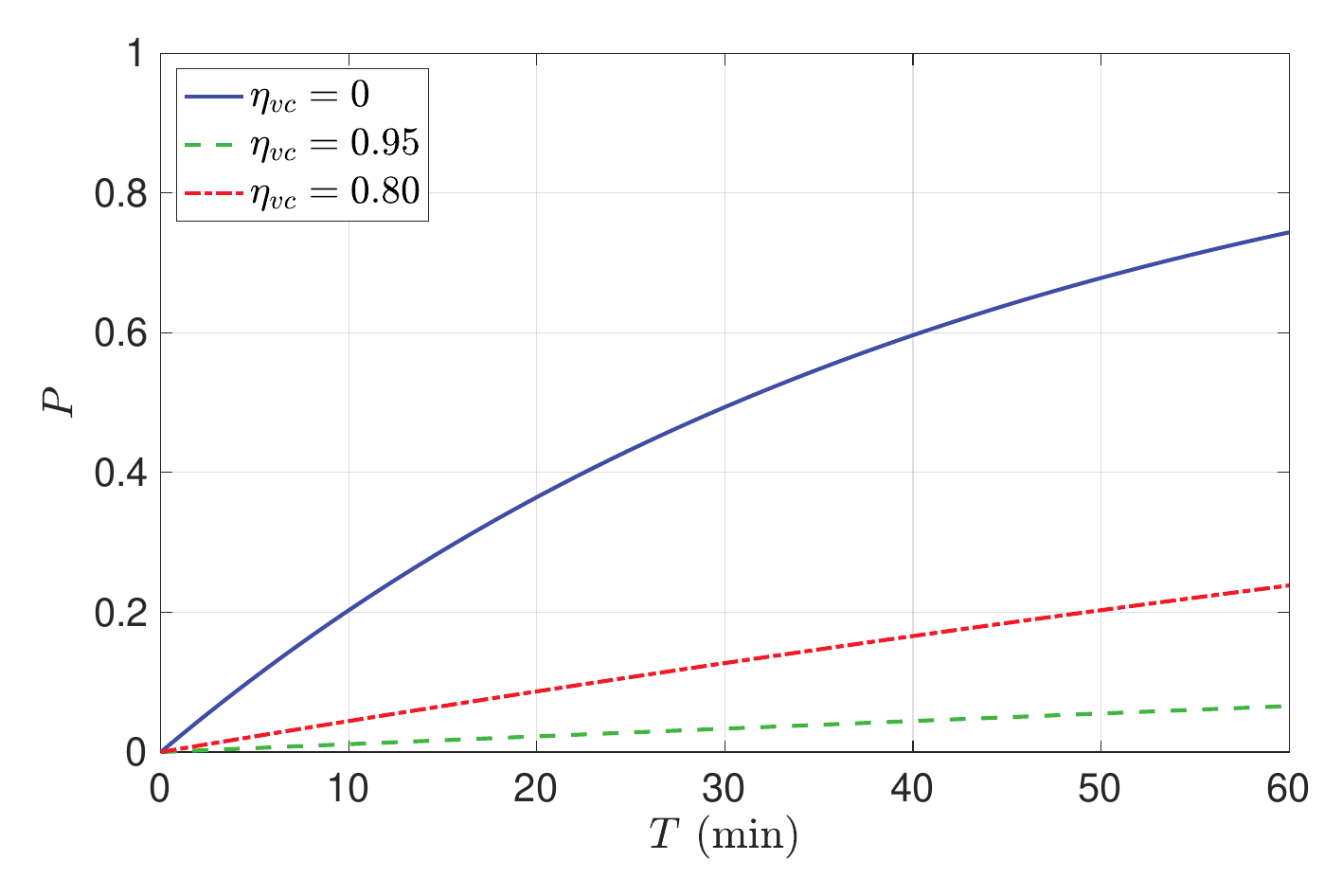}}
	\caption{(a) comparison of infection probability over time of different strains of SARS-CoV2. (b) Comparison of infection probability of vaccinated and unvaccinated cases.  }
	\label{fig:RH50-variant}
\end{figure}

\begin{figure}[!t]
	\centering
	\subfigure{
		\includegraphics[width=0.48\textwidth]{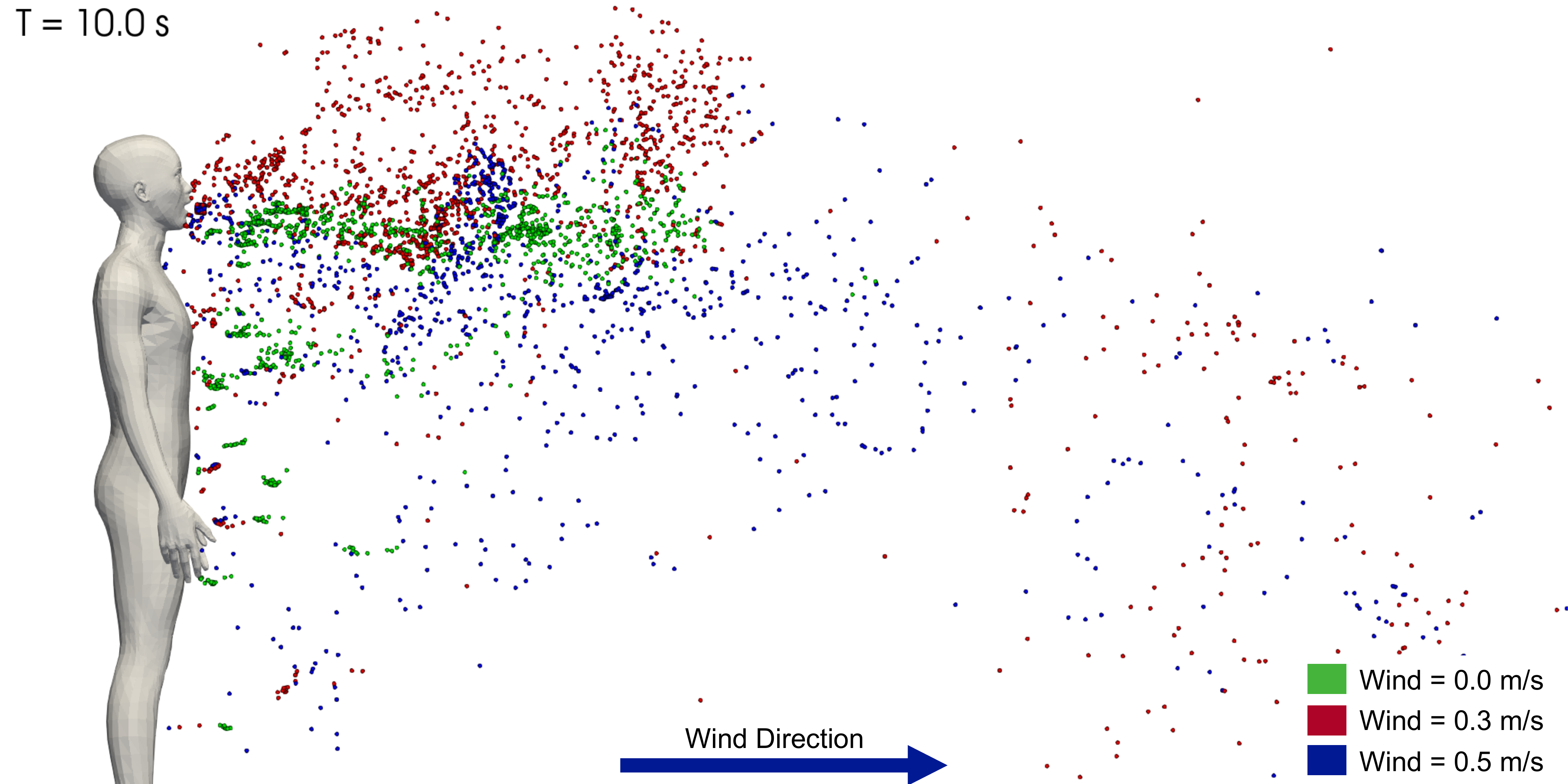}    }
	\subfigure{ 
		\includegraphics[width=0.48\textwidth]{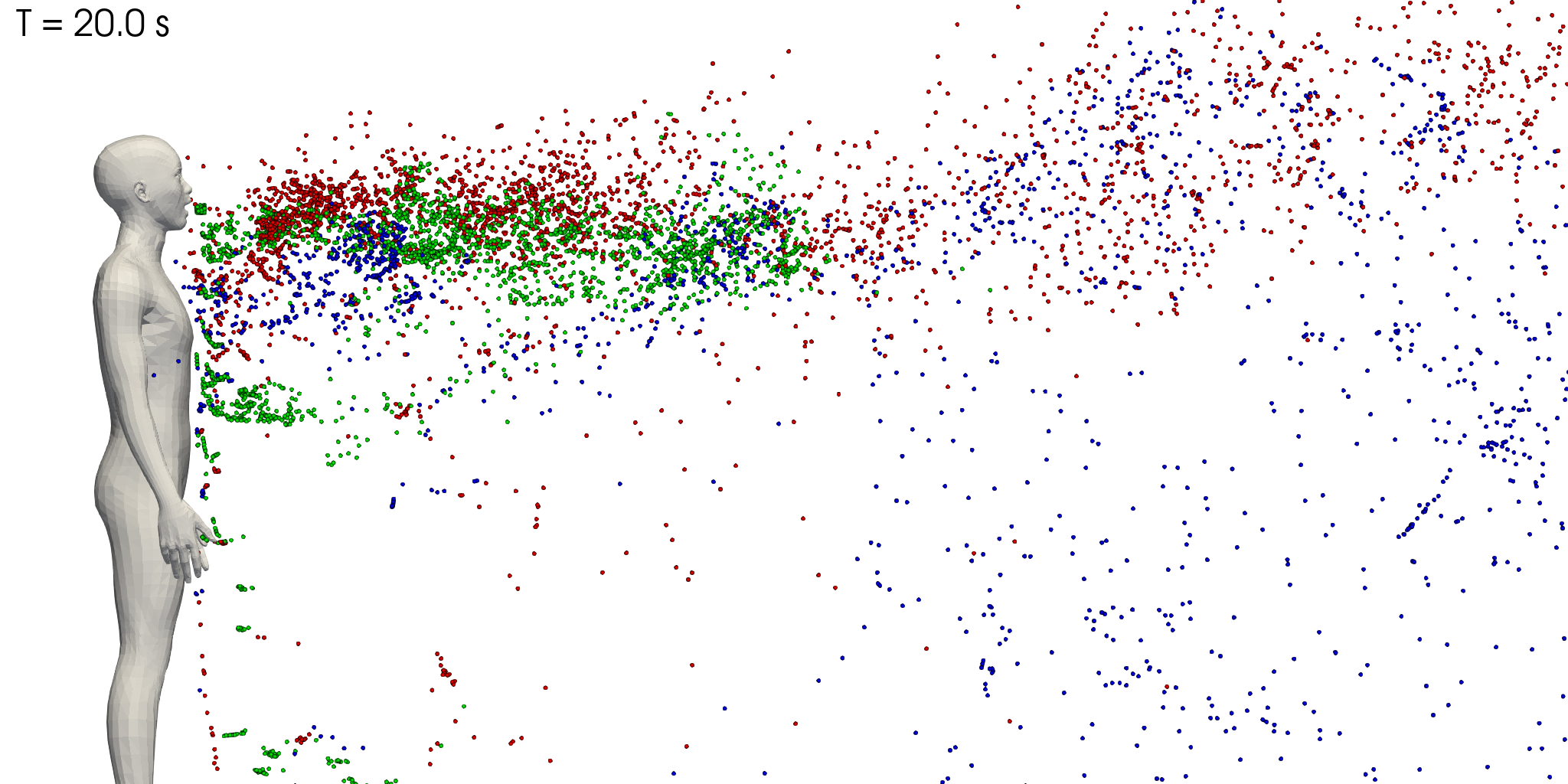}    }
	\subfigure{ 
		\includegraphics[width=0.48\textwidth]{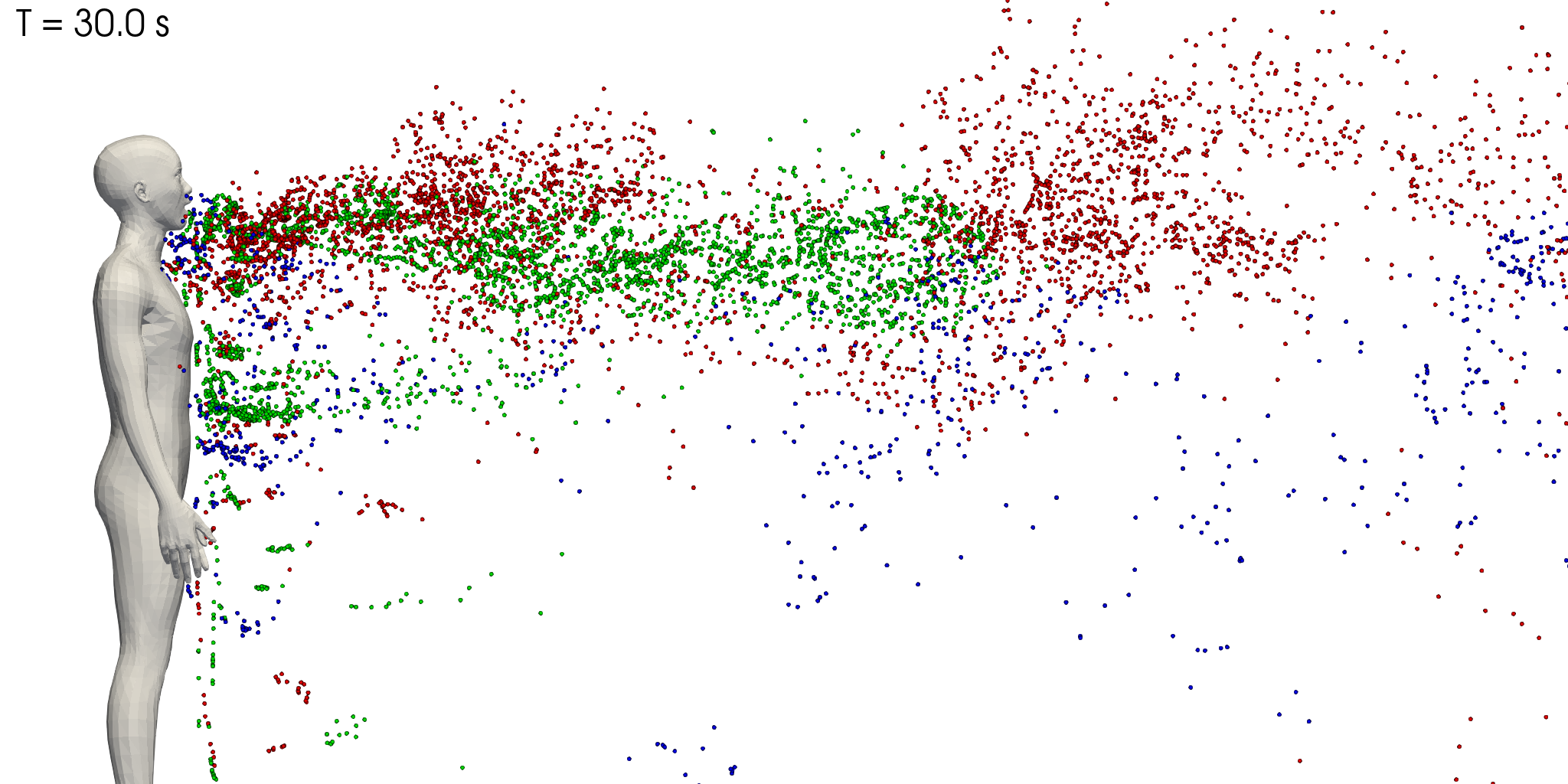}    }
	\subfigure{ 
		\includegraphics[width=0.48\textwidth]{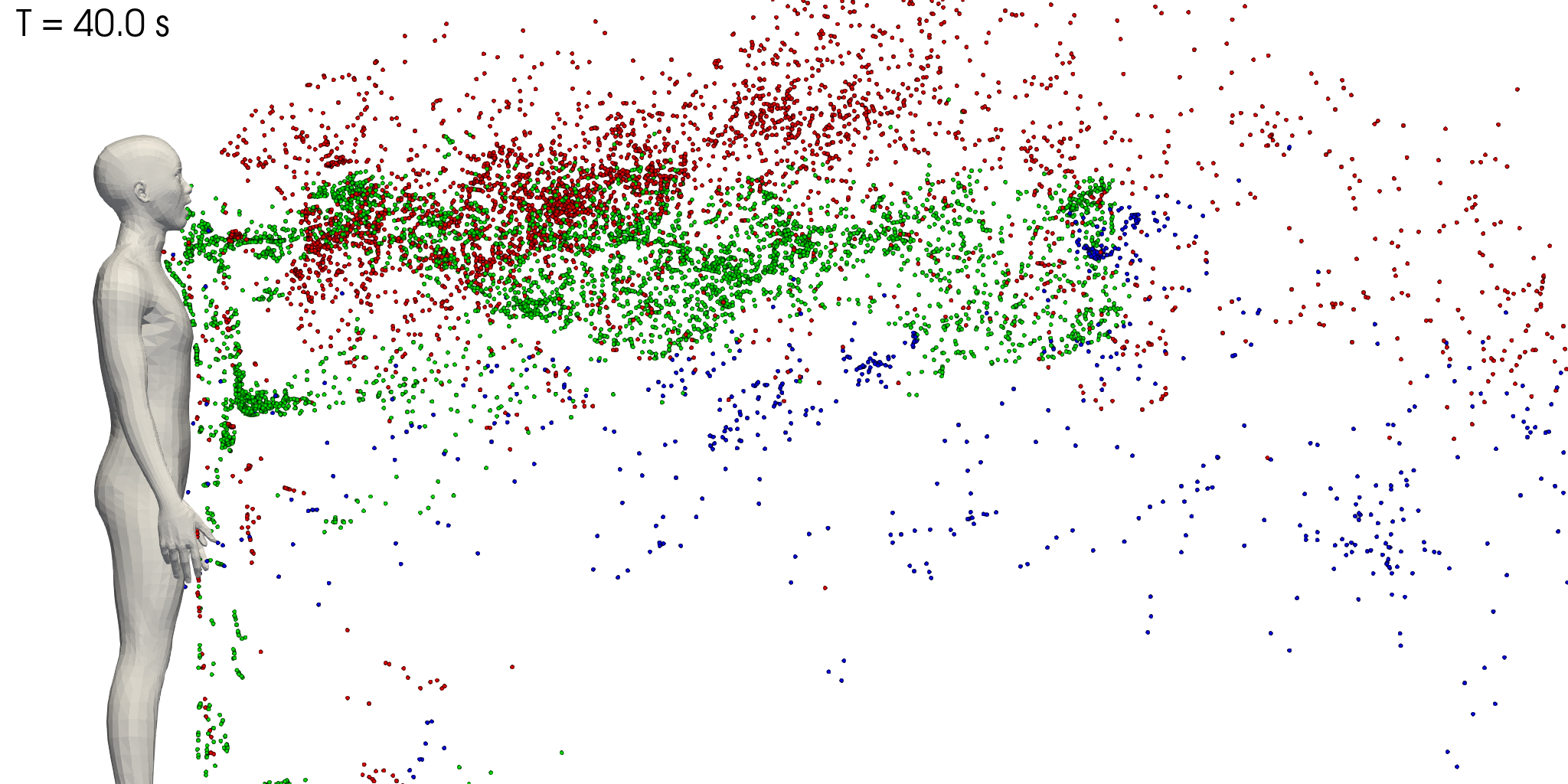}    }	
	\caption{ Comparison of droplet dispersion of the quiescent and the rear wind cases at different time instants. }
	\label{fig:wind effect}
\end{figure}

\begin{figure}[!t]
	\centering
	\includegraphics[width=0.52\textwidth]{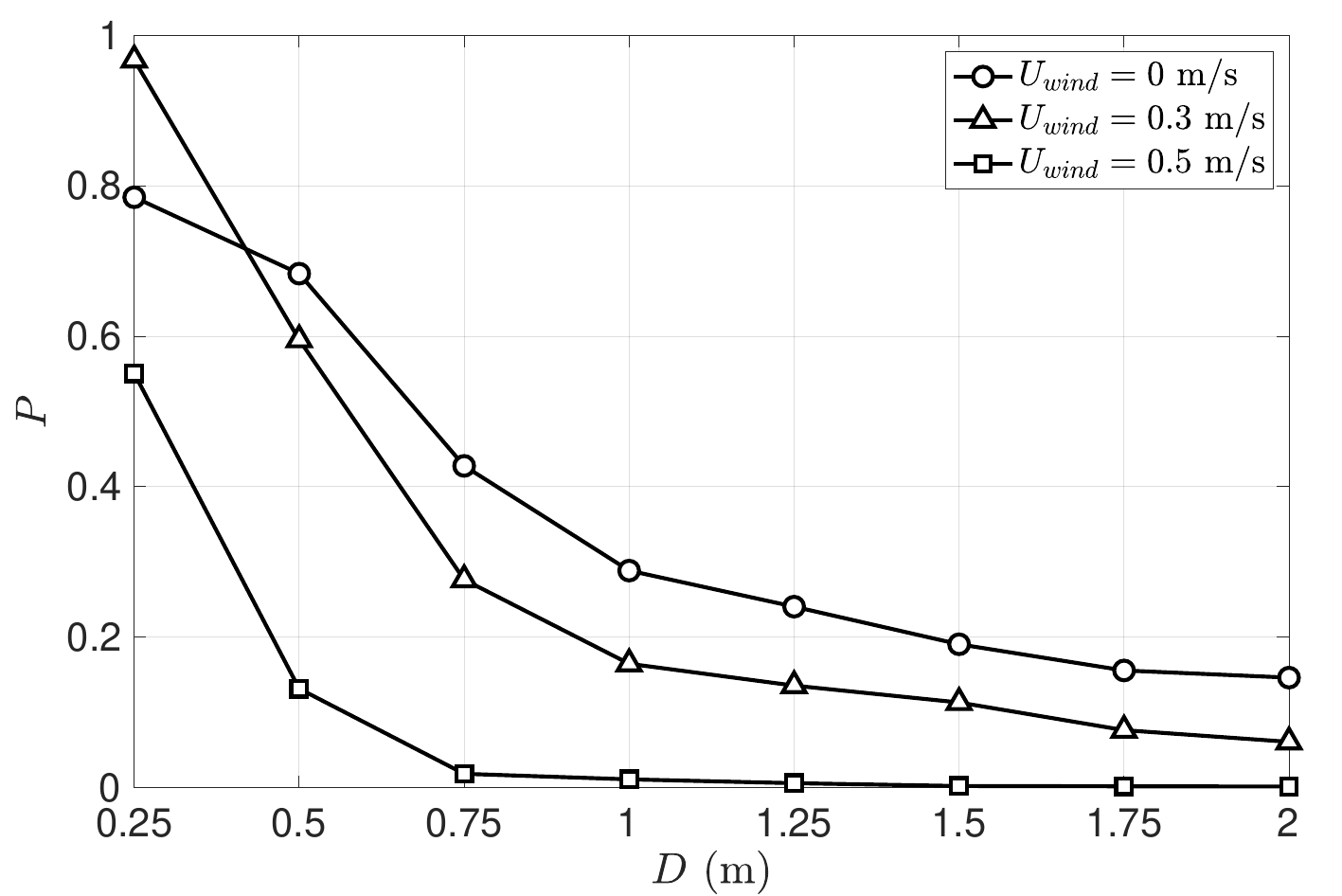}    
	\caption{ Probability of infection after 15 min. of exposure under different wind conditions.}
	\label{fig:wind effect_p}
\end{figure}

  {\textbf{Effect of wind}: Given the uncertainties involved in determining the parameters of the dose-response model, its usefulness in predicting absolute values of infection risk is limited.  However, the dose-response model is useful in providing a relative measure of the infection risk under different environmental conditions such as ambient airflow, humidity, temperature, etc. assuming that the uncertainty of the modeling parameters is not significantly influenced by environmental conditions. In this section we investigate the role of ambient airflow in altering the infection risk. In addition to the quiescent condition, we consider two cases of ambient wind blowing in the direction of the speech flow at $ 0.3 $ m/s (1.08 km/hr) and $ 0.5 $ m/s (1.8 km/hr) from behind the infected person. A comparison of the droplet dispersion at four time instants is shown in Fig.~\ref{fig:wind effect}. The droplets from the quiescent case are colored in green, while the 0.3 m/s and 0.5 m/s rear wind cases are colored in red and blue, respectively. The droplets from the three cases are overlaid on the same figure for each time instant for ease of comparison. The degree of accumulation of droplets in front of the infected subject seen in the quiescent case is greater than in the rear wind cases. As a result of the wind, the droplets are carried and dispersed by the airflow.  The unsteady flow, due to the interaction between wind and infected subject, results in greater dispersion of the droplets in the vertical direction. However, the level of dispersion depends on the wind speed. The dispersion due to 0.5 m/s wind is greater than the dispersion due to 0.3m/s wind.  It is interesting to note that the local concentration of droplets due to the 0.3m/s wind, within 0.5m from the infected person, is as high as the quiescent case. However, for distances greater than 0.5m, the dispersion of 0.3 m/s cases is greater than the quiescent case but less than the 0.5 m/s cases.  This trend can be observed in a quantitative form in Fig.~\ref{fig:wind effect_p} in which the infection risk of the two wind cases and the quiescent case are compared. The infection risk at any distance, except $ D=0.25 $m, progressively decreases with the wind speed.  For distance greater than 0.5m, the infection risk when the wind speed is 0.3 m/s on average is decreased by half compared to the quiescent case. Upon increasing the wind speed to 0.5m/s the infection significantly decreases for distances larger than 1m.  The dependence of the infection risk on distance shows similar trends for all three cases. At $ d=0.25 $m, the infection risk of the 0.3 m/s wind case is higher than the no-wind case. This counterintuitive result can be understood by accounting for the interaction between the wind and natural convection plume from the human body. The upward fluid flow emanating around the infected subject is shifted in the direction of the wind, as a result of which the plume carries come of the falling droplets upwards increasing the number of droplets immediately in front of the infected subject. In contrast, the plume from the infected subject does not significantly interact with the droplets. Due to the greater accumulation of droplets at short distances in the 0.3 m/s wind case, the infection risk in this region is greater than in the no-wind case.  
	
	It is important to note that the inclusion of an at-risk human subject in front of the infected person in the numerical simulation could significantly alter the flow and how the droplets enter the breathing zone, thereby altering the infection risk.  The direction of the wind would also significantly alter the droplet dispersion and consequently the infection risk. Specifically, for the wind blowing opposite to the speech jet, wind speeds greater than 0.2m/s could prevent the droplets from being dispersed to a distance greater than 1m. This is because the velocity of the speech rapidly decays to less than 0.2m/s within 1m from the infected subject (see Fig~\ref{fig:mesh_study}). Therefore, the infection risk for distances greater than 1m could be nil.}

\textbf{Effect of variants and vaccines}: The probability of infection can change significantly due to variant strains of SARS-CoV-2 and vaccination. To account for such factors a generalized form of the equation for infection probability was introduced through Eq.~\ref{eqn:Prisk-vaccine}. By altering the parameter $ \alpha $, the equation can be adapted for variant strains of higher transmissibility factor $ \tau $, and vaccinated individuals. The transmissibility of B.1.1.7 strain (alpha variant) and the B.1.617.2 strain (delta variant) is $ 29\% $ and $ 145\% $ higher than the original strain, i.e. $ \tau = 1.29 $ and $ \tau=2.45 $ for the alpha and the delta variant, respectively. Evidently, the value $ \tau = 1 $ corresponds to the original strain. A comparison of how the $ P $, evaluated at a distance of 1 m, increases with time for the original strain, the alpha and the delta variants is plotted in Fig.~\ref{fig:RH50-variant}a.  For the duration of exposure of less than 20 min, we find that there is a significant difference between the risk of infection of the original strain and the delta strain. This difference gradually reduces over time as P saturates to its maximum value. The probability of infection for the delta variant is approximately 2 times greater than the original strain.  However, the difference between the infection probability of the alpha and the original strain is not very significant. 

Through the risk evaluation model presented in this work, it is also possible to incorporate the effect of vaccination on the probability of infection. As discussed in the previous section, the generalized form of Eq.~\ref{eqn:Prisk-vaccine} can be applied to vaccinated cases by using the expression $ \alpha = 1-\eta_{vc} $. For unvaccinated cases, we can set $ \eta_{vc}=0 $. To evaluate how the infection risk changes due to vaccination, we choose two values for $ \eta_{vc} $, $ 80\% $ and $ 95\% $, which approximately correspond to the efficacies of the AstraZeneca and Pfizer vaccines, respectively. The comparison of $ P $ evaluated at a distance of 1m from the infected person for the different values of $ \eta_{vc} $ is presented in Fig.~\ref{fig:RH50-variant}b. The infection probability remains below $ 20\% $ for exposure periods of up to 40 mins for $ \eta_{vc} = 0.85 $ and it remains well below $ 10\% $ for exposure period plotted in the figure.

\begin{figure}[!tb]
	\centering
	\subfigure[]{
		\includegraphics[width=0.48\textwidth]{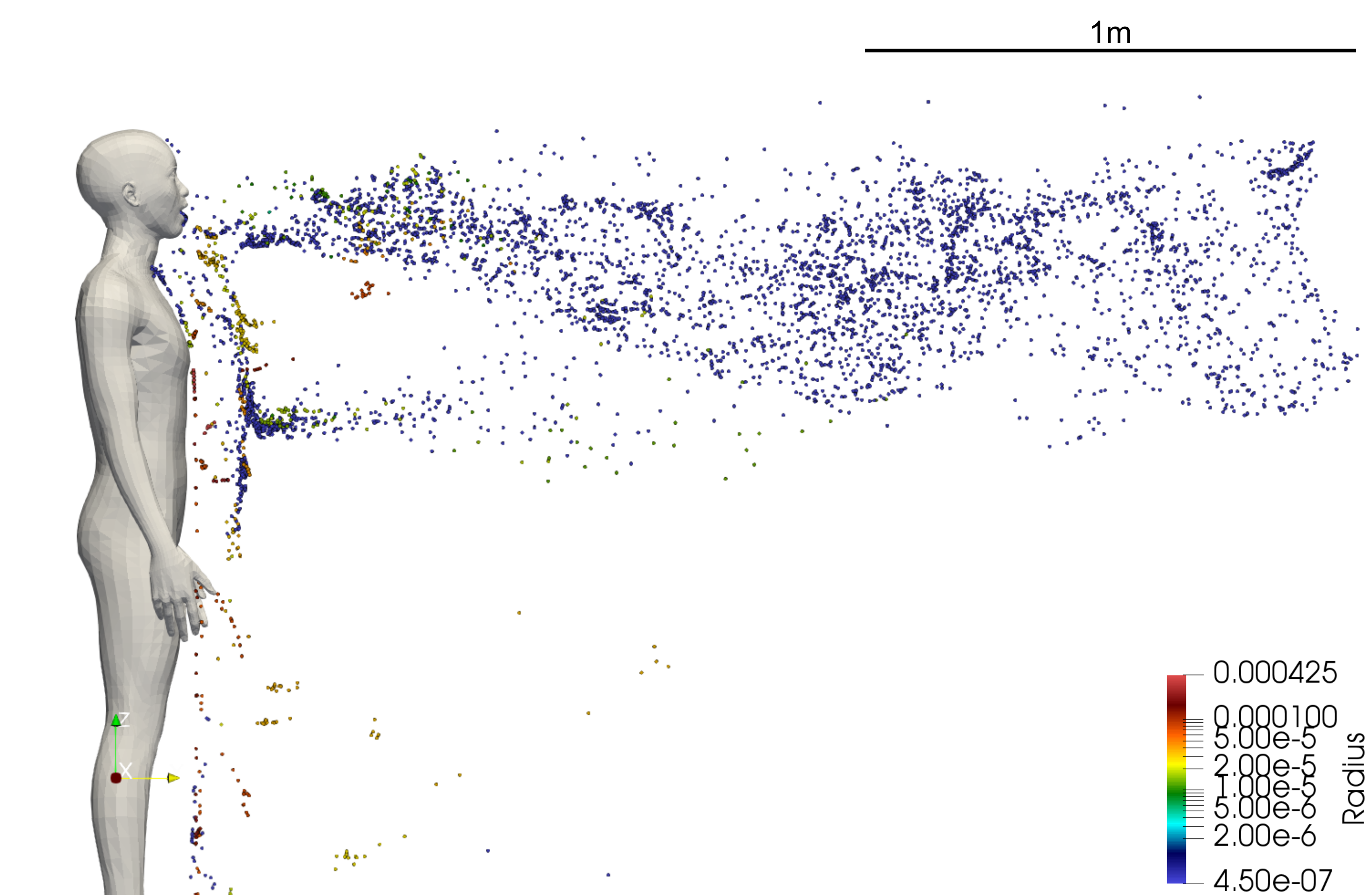}    }
	\subfigure[]{ 
		\includegraphics[width=0.48\textwidth]{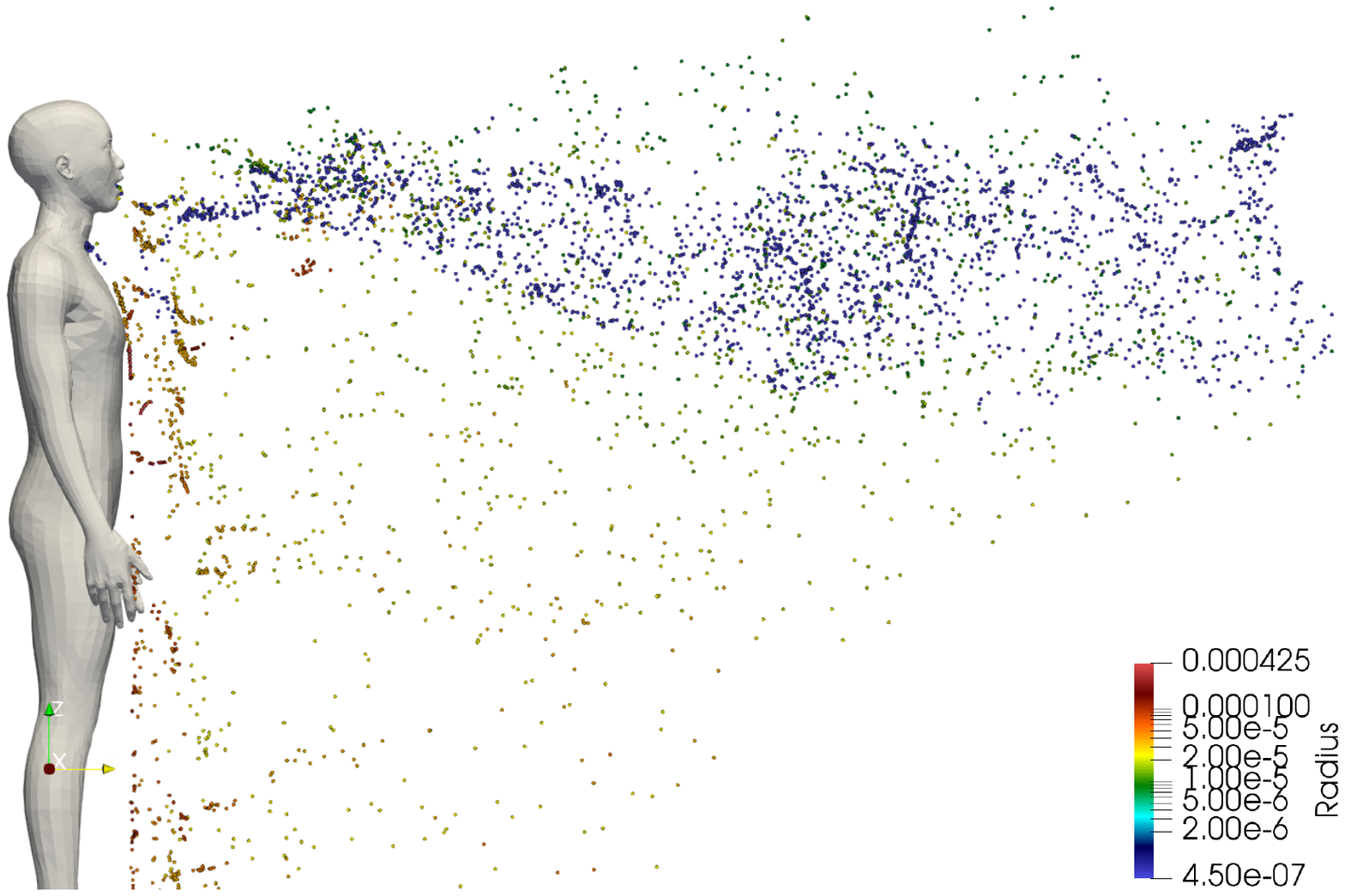}    }
	\caption{A snapshot of droplet dispersion at $ t=24 $ s for (a) $ RH=50\% $ (b) $ RH=100\% $.}
	\label{fig:droplet-viz-RH}
\end{figure}

\begin{figure}[!tb]
	\centering
	\includegraphics[width=0.52\textwidth]{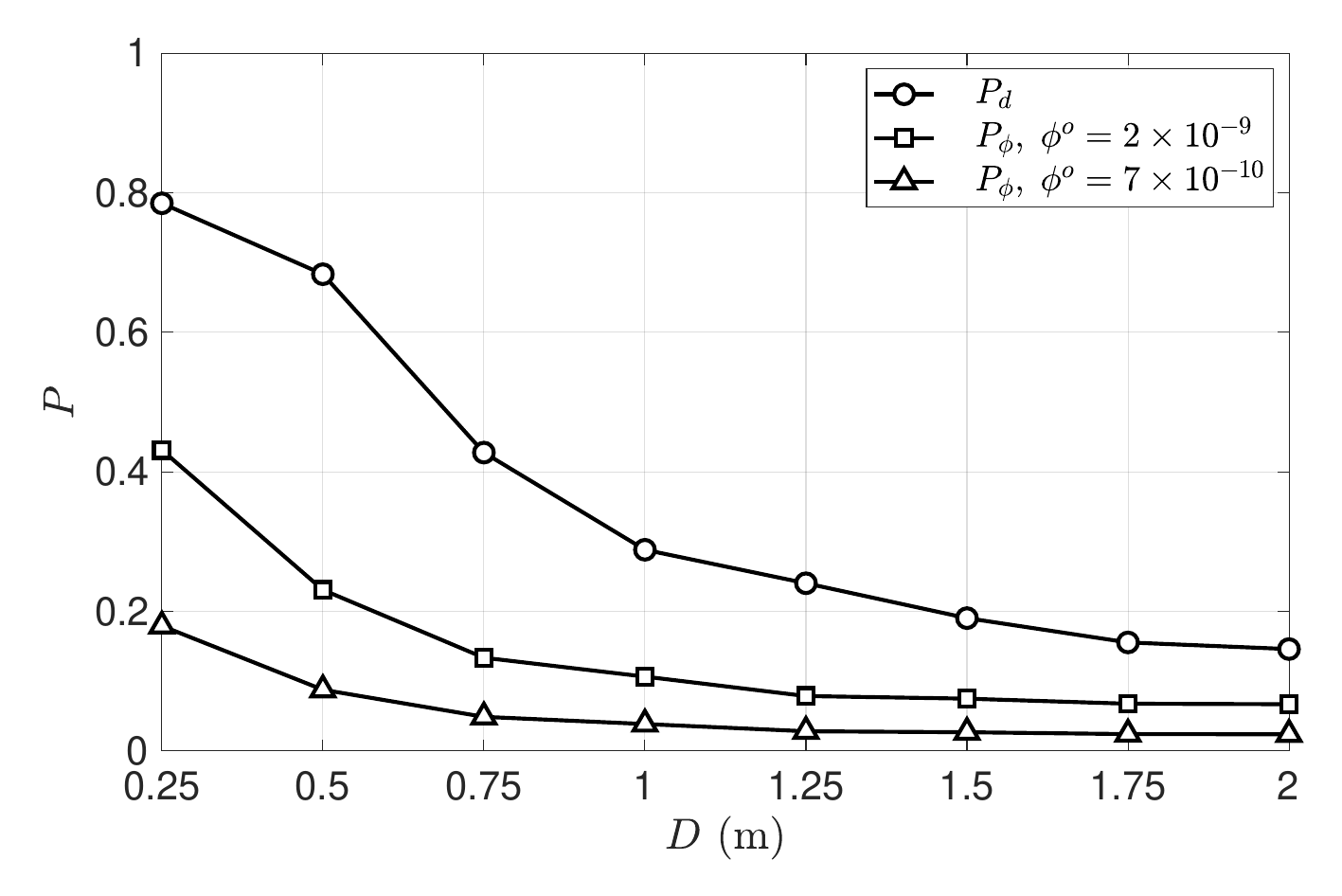}    
	\caption{Comparison of probability of infection over distance based on $ P_d $ and $ P_{\phi}$ for an exposure period of 15 mins.}
	\label{fig:scalar_p}
\end{figure}

  { \textbf{Well-mixed approximation based infection risk}: Next, we compare the infection prediction based on the discrete droplet-based approach of the present work with that from the well-mixed approximation. The number of virions as a function of space and time presented in Eq.~\ref{eqn-N_phi_b} depends on droplet volume fraction at source $ \phi^o $ which has been reported to be in the range $ 2\times 10^{-9} - 1 \times 10^{-8} $ by Leung et al.\cite{leung2020}  The value for $ \phi^o $ can also be arrived at based on the droplet diameter distribution, rate of droplet injection, and volume of flow exhaled in the present speech model. Assuming that all the droplets less than 25 micrometres in diameter get aerosolised within a short period after being ejected from the mouth, we get $ \phi^o=6.99\times 10^{-10}\approx 7\times 10^{-10}$. This value is about 1/3 of the corresponding minimum value reported in the literature ($ 2\times 10^{-9}$), which implies that the droplet diameter distribution adopted in this work under-predicts the measured aerosol generation during speech. Nevertheless, it can still be informative in the study of the effect of intervention strategies, humidity, vaccines, etc. For comparing the present approach of using discrete droplet simulations and passive scalar based well-mixed approximation, $  \phi^o= 7\times 10^{-10} $ is the appropriate choice. A comparison of the probability of infection from the discrete droplet-based estimation ($ P_d $) and the passive scalar based estimation ($ P_{\phi} $) is presented in Fig.~\ref{fig:scalar_p}. The passive scalar based estimation significantly underpredicts infection risk when compared to the discrete droplet-based estimation. The ability of a well-mixed approximation to accurately predict the local droplet volume fraction depends on the magnitude of the volume fraction at the source. It is reasonable to expect the approach to be more accurate for large values of $ \phi^o $ at source, conversely, less accurate for small values of $ \phi^o $. For droplet generation during a speech, the value of $ \phi^o $ is perhaps not sufficient leading to a disparity in the infection risk compared to the discrete droplet-based evaluation. Also plotted in Fig.~\ref{fig:scalar_p} is $ P_{\phi} $ corresponding to  $ \phi^o = 2\times 10^{-9} $ (measured data \cite{leung2020}) to contrast against that corresponding to $  \phi^o= 7\times 10^{-10} $.  Even with larger $ \phi^o $ the infection risk is significantly lower than the discrete droplet-based infection risk. 
	
	The method of estimation $ P_\phi $ uses only the local values of $ Y_s $ at the nose of the susceptible subject. This approach over predicts the number of droplets inhaled. This is because the scalar concentration is maximum at the centre of the speaking flow jet and decays away from the centre at any given distance from the source of the jet. A method of approximating the mean value of $ Y_s $ based on a top-hat and a Gaussian approximation can be found in the work of Singhal et al.\cite{singhal2021}. The top-hat approximation yields a pre-factor of 0.462 to $ Y_s $ at the centre of the jet, while the Gaussian approximation yields a corrected but lower $ Y_s $ that is a function of distance from the source of the jet. Adopting either the top hat or the Gaussian approximation would result in a lower $ P_{\phi} $ compared to the ones presented in Fig.~\ref{fig:scalar_p}. }

\subsection{Role of humidity}
Direct simulation of droplet dispersion, as opposed to simulations of scalar transport as a proxy for aerosols, enables the investigation of the effect of environmental factors like temperature and humidity on droplet aerosolization and dispersion, and consequently on the risk of infection. We extend the numerical simulation of droplet dispersion during speech presented in the previous section to study the effect of humidity on infection risk. The relative humidity of the ambient environment could significantly affect the evaporation of medium and large droplets. This leads to the prevention or aerosolization of medium droplets that have a higher virion count compared to that of smaller droplets. As a consequence, the concentration of virions could be significantly altered by humidity, thereby influencing the probability of infection. In order to investigate role of humidity on infection risk, we carried out a numerical simulation of droplet dispersion during speech under three different humidity conditions $ 10\%, 50\% $ and $ 100\% $. The numerical setup and the boundary conditions are identical to those of the simulation in the previous section.  The only parameter varied is the relative humidity ($ RH $). Three separate simulations were carried out in which the relative humidity was set to $ 10\%, 50\% $ and $ 100\% $. 

\begin{figure}[!t]
	\centering
	\subfigure[]{
		\includegraphics[width=0.48\textwidth]{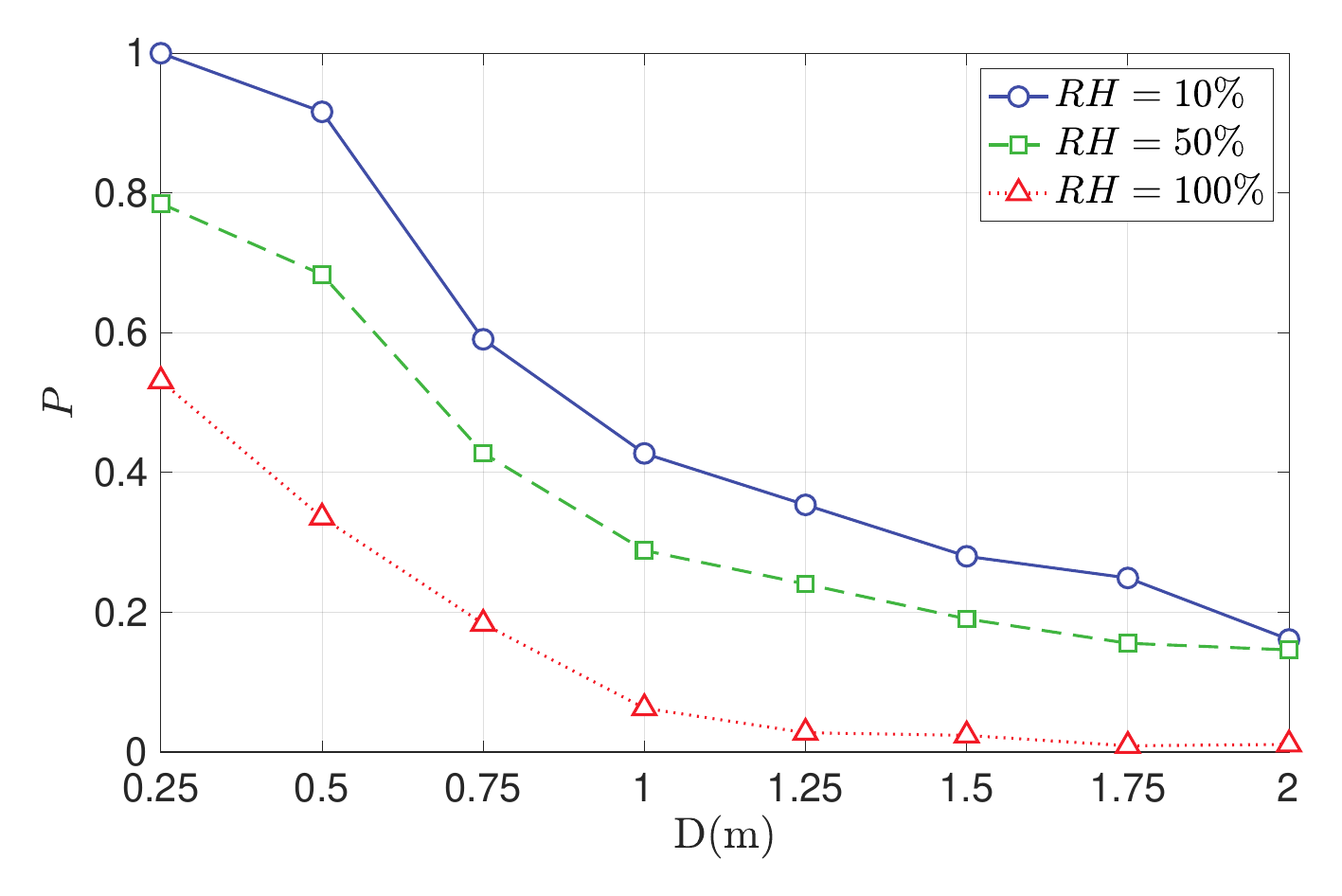}    }
	\subfigure[]{ 
		\includegraphics[width=0.48\textwidth]{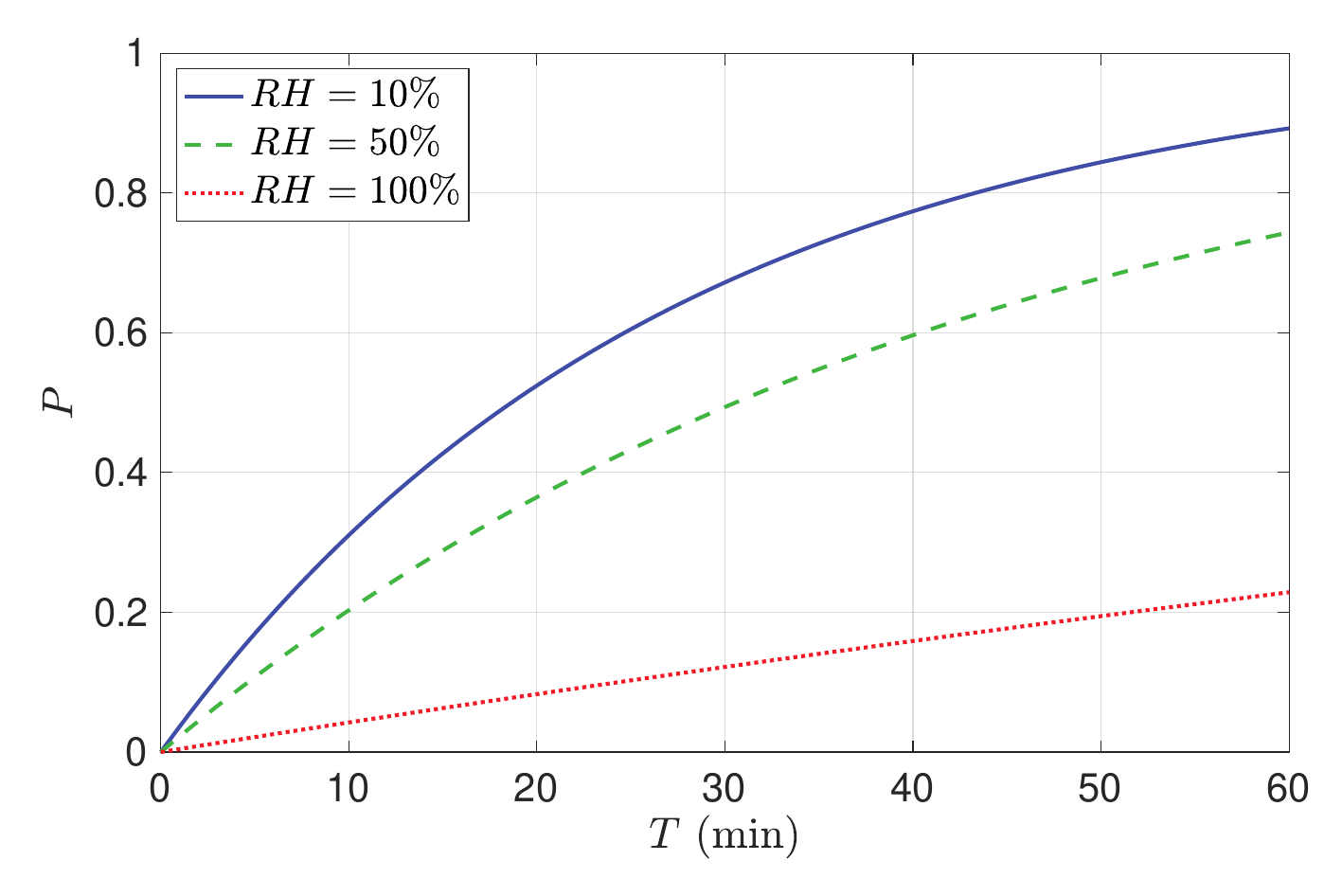}    }
	\caption{(a) Variation of infection probability over (a) distance  from the infection source, and (b) time of exposure under different humidity conditions}
	\label{fig:Rh-effect-dist-time}
\end{figure}

In Fig.~\ref{fig:droplet-viz-RH}, a snapshot of the droplet dispersion at $ t=24 $ s for $ RH=50\% $ and $ RH=100\% $ is presented.  For the $ RH=100\% $ case, lack of evaporation prevents the aerosolization of medium-sized droplets. Droplets smaller than $ 5\;\mu$m remain airborne for prolonged periods. The velocity of some of the droplets between the sizes of 5 and $ 50\;\mu$, whose timing in ejection matches the peak velocity of the speaking model, is initially dominated by the fluid velocity. However, as the flow velocity decreases away from the mouth due to dissipation, the influence of gravity dominates the velocity of these droplets. This results in the droplets settling to the ground under the influence of gravity. In contrast, at lower $ RH $ values, the medium droplets in question can get partially or completely aerosolized and remain airborne.  As a result, the concentration of aerosols and consequently the virions depends directly on the relative humidity of the ambient environment. This effect is quantified through the evaluation of the infection probability which directly depends on the local virion concentration. The variation of infection probability over distance from the infection source of the different humidity cases are compared in Fig.~\ref{fig:Rh-effect-dist-time}a. As the local droplet concentration is strongly influenced by the evaporation of medium droplets with higher virion count,  the number of virions likely to be inhaled decreases with increasing humidity. The consequence of which can be seen in Fig.~\ref{fig:Rh-effect-dist-time}a and Fig.~\ref{fig:Rh-effect-dist-time}b. The trend of $ P $ decreasing with distance is consistent across all the humidity cases. However, for a given distance from the infection source, the probability of infection is lower for higher humidity.  Fig.~\ref{fig:Rh-effect-dist-time}b plots the evolution of $ P $ over time evaluated at a distance of 1 m from the infection source for the different humidity cases. From these plots, it could be inferred that for a given temperature higher humidity lowers the risk of infection. However, the magnitude of reduction in the probability of infection could be strongly influenced by the temperature of the ambient environment. 

\section{Summary and Discussion}
In this work, we have developed a framework for quantifying the risk of infection due to airborne diseases like COVID-19 using the dose-response model in droplet-dispersion simulations. We have detailed in this work a methodology for estimating the inhalation dose, by measuring the volume of droplets/aerosols (proxies for virion count) in the breathing zone, directly from droplet dispersion simulations. We have adopted this framework to estimate the risk of infection from a person ejecting droplets while speaking. For this, we have carried out a numerical simulation of droplet dispersion from a single person standing and speaking continuously in an isolated environment. The infection probability was found to decrease with distance from the infected person. The magnitude of infection probability is strongly influenced by the minimum infection dose $ N_0 $ which can be seen in Fig~\ref{fig:RH50-PvsDist}a. If the minimum infection dose is assumed to be small then the infection risk remains relatively large even at distances as far as 2m from the infected person. On the other hand, if the infection dose is large then the infection probability is small even at a distance of 1 m.   {The role of wind blow from the behind the infected person towards the at-risk subject were considered and found to lower the risk of infection compared to the poorly ventilated case due to greater droplet dispersion caused by the wind.} This prediction can drastically change depending on the ambient environment's flow conditions.  Therefore, we would not like to make any specific recommendations on social distancing based on these results. The main purpose of this work is to demonstrate the applicability of the framework of risk estimation using the dose-response model in droplet dispersion simulations. 
\  {A quantitative comparison of the well-mixed approach and the present approach was considered in this work. The comparative analysis showed that the well-mixed approach significantly under predicts the infection risk for the same simulation conditions. However, the dependence of the infection risk on distance from the infected subject was found to be similar for the well-mixed and the present approach.}

A generalized form of the dose-response model that can incorporate the effect of increased transmissibility of various strains of COVID-19 and the effect of vaccination has been presented in this work.  A comparison of the infection risk due to variant strains of higher transmissibility was with the standard strain was presented. The results show that for exposure duration the infection risk of variant strain can be significantly higher than the standard strain. The difference in infection between strains decreases as exposure duration increases. Similarly, we also presented a comparison of the infection risk for a vaccinated person with that for an unvaccinated person. The probability of infection for vaccinated persons is very low for short and medium exposure duration. However, the infection risk can increase significantly provided the exposure duration is very long. 

One of the main advantages of droplet dispersion simulations is their ability to investigate the effect of environmental factors, keeping all other variables fixed, such as temperature and humidity on the infection risk. To demonstrate this point, an investigation of the effect of humidity of the ambient environment on the infection risk was carried out. This is an aspect that cannot be analyzed using well-mixed room average analysis or passive scale based aerosol transport models. The results of our analysis show that the infection risk is strongly influenced by humidity due to its effect on the evaporation of medium and small droplets. The infection risk at a given distance from the infected person has an inverse dependence on humidity. Lowering the ambient humidity increases the risk and vice-versa. This relationship between infection risk and humidity depends, to some extent, on the droplet diameter distribution adopted in the simulation. For example, a hypothetical droplet diameter profile that includes only aerosols may not result in a strong dependence of infection risk on humidity as the results of this study indicate.

\section*{Acknowledgements}
This work was supported by JST CREST, Grant Number JPMJCR20H7, Japan, and through the HPCI System Research Project (Project ID:  hp210086). R.B acknowledges the support for this work by JSPS KAKENHI Grant Number 22K10596.

\section*{Author contributions statement}
RB, AI, MT developed the dose-response model. RB, MT, MY developed the speaking model. RB, CGL, MT developed flow solver. RB carried out the simulations. RB, MT analyzed the data. RB wrote the manuscript. All authors reviewed the manuscript and provided inputs to the final edit.

	\bibliography{references.bib}
\appendix
\renewcommand{\thesection}{\Alph{section}}
\renewcommand{\thesubsection}{\Alph{section}.\arabic{subsection}}
	\section{Appendix}
\subsection{Grid Convergence}

\begin{figure}[!t]
	\centering
	\subfigure[]{
		\includegraphics[width=0.45\textwidth]{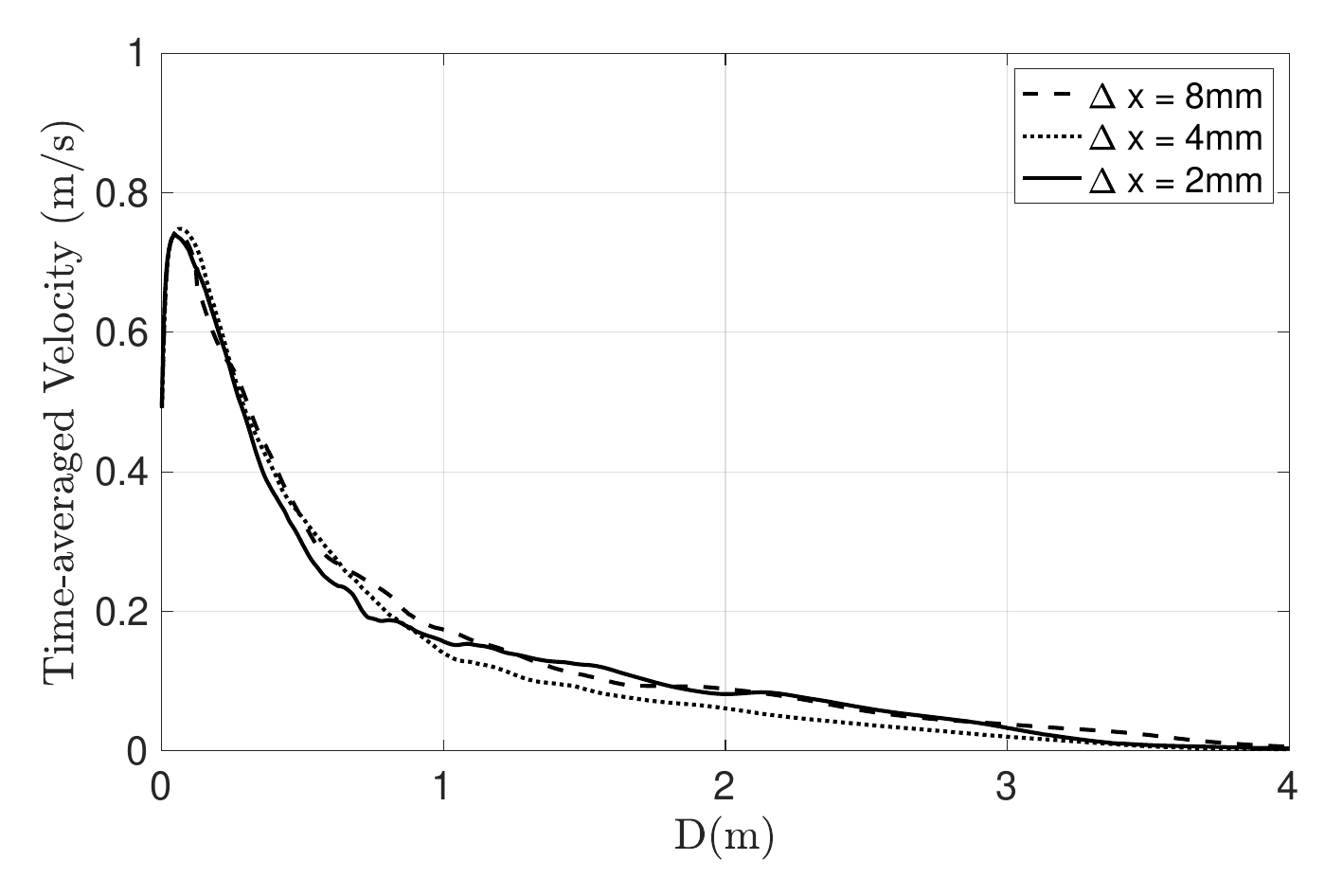}}    
	\subfigure[]{
		\includegraphics[width=0.45\textwidth]{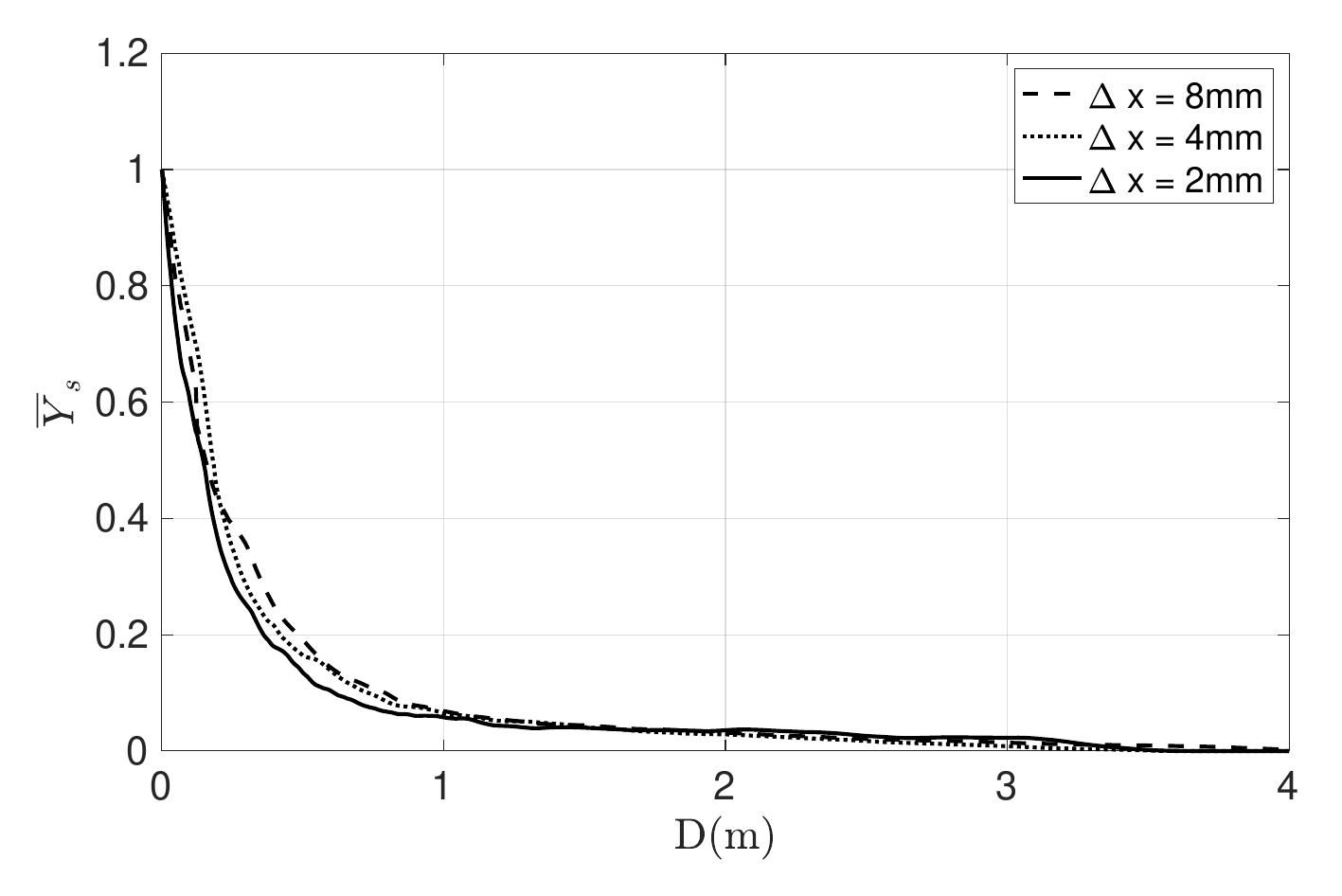}}
	\caption{(a) Comparison of axial velocity along the axial center line of the speaking jet flow for the coarse, medium and fine meshes. (a) Same as (a) with the comparison of the scalar concentration. }
	\label{fig:mesh_study}
\end{figure}

The numerical simulation of the speaking flow was carried out with three different meshes in which the mesh spacing in the a region of at least $ 1m \times 0.5m \times 0.5m $ along the axial, lateral and vertical directions, respectively, in front of the mouth geometry from which the speaking flow emanates. The mesh spacing of the three meshes in this region were 8mm (coarse), 4mm (medium) and 2mm (fine), respectively. As the distance from the mouth increases the mesh spacing progresively decreases for any given mesh. For example, for the medium mesh, the mesh spacing increase from 4mm to 8mm, then from 8mm to 16mm and so on as the cube sizes are changed (see Fig.~3). The time averaged axial velocity along the center line of the speaking jet flow is plotted for the three meshes in Fig.~\ref{fig:mesh_study}a. The simulations were carried out for a physical time of 80s for three cases and the time averaging of the data was done for the same period. We find that there is no significant variation between the three results. Therefore we proceed adopt the fine mesh for the present study.   In Fig.~\ref{fig:mesh_study}b, a comparison of the scalar concentration is presented for the three meshes. The scalar concentration is derived from the mass fraction of the vapor phase of the liquid droplets($ Y_d $), which is defined as follows
\begin{equation}
	\overline{Y}_s = \frac{\overline{Y}_d-Y_{d}^{\infty}}{{Y}_{d}^{o}-Y_{d}^{\infty}}
\end{equation}
where the overbar indicates a time-averaged quantity,  $ Y_{d}^{\infty}  $ and $ Y_{d}^{o}  $  are the ambient mass fraction of the vapor phase of the droplet and that at the source of the injection. 

\subsection{Solver framework and simulation environment}
A multi-physics solver known as CUBE\cite{jans18,nishiguchi19} has been used for all the numerical simulations presented in this work. CUBE is a finite volume solver based on a hierarchical meshing framework known as the building cube method (BCM)\cite{naka03}. The meshing framework allows local mesh refinement enabling the high resolution in regions of interest while limiting the overall cell count. The supercomputer Fugaku has been used for carrying out the numerical simulation presented in this work. Fugaku comprises 158,976 nodes. Each node is equipped with a Fujitsu A64FX processor, which consists of 48 compute cores and 4 additional cores, and a memory of 32 GiB. The nodes are interconnected with 28Gbps, 2 lanes, 10 port TofuD interconnect.

\subsection{Infection probability post vaccination}
\label{asec:vacc}
The probability of infection can be written as 
\begin{equation}
	P = 1- e^{(-\frac{N(x,t)}{N_0})} = 1- \exp^{(-\frac{N_c }{N_0}t)}
	\label{eqn-A1}	
\end{equation}
where $ N_c $ is the virion dose per unit time under steady state conditions of exposure such as speaking, singing etc. The exact form of $ N_c $ is not pertinent to the present derivation, it can deduced from Eq.~6 of the main manuscript. The probability of no infection can be defined as $ \hat{P} = 1-P$. The net probability of no infection is given by 
\begin{align}
	\hat{P}_{net} & = \int_{0}^{\infty}  \exp^{(-\frac{N_c }{N_0}t)} dt \nonumber \\
	\hat{P}_{net} & = \frac{N_0}{N_c }
\end{align} 
Assuming that the vaccination leads to reduction in infection probability due to increase in $ N_0 $, the net probability of no infection for a vaccinated individual can be expressed as 
\begin{equation}
	\hat{P}^{vc}_{net}  = \frac{N_{0}^{vc}}{N_c }
\end{equation}
Let $ \eta_{vc} $ be the efficacy of a given vaccine. The probability of no infection with vaccination decreases by a factor of $ \hat{P}^{vc}_{net}\eta_{vc} $ if the vaccination is not administered. Therefore, we can express a relationship between $ \hat{P}_{net} $ and $ \hat{P}^{vc}_{net} $  as follows
\begin{equation}
	\hat{P}_{net} = \hat{P}^{vc}_{net} - \hat{P}^{vc}_{net} \eta_{vc}.
\end{equation}
Rearranging the terms we get
\begin{equation}
	\hat{P}^{vc}_{net} = \frac{\hat{P}_{net}}{1-\eta_{vc}}.
\end{equation}
Assuming an exposure to the same average virion dose per unit time, we can obtain an expression for $ N_{0}^{vc} $ from the above two equation as

\begin{equation}
	N_{0}^{vc} = \frac{N_{0}}{1-\eta_{vc}}
\end{equation}

The interpretation of the above two equations is straightforward. If the vaccine efficacy the $ 100\% $ then the minimum number of virions needed for infection will be $ \infty $. Consequently, $ \hat{P}=1 $ at all time instants $ t $ and $ P = 0 $ and  the net probability of no infection will be $ \infty $. On the other hand, if the vaccine efficacy is 0 then the $ 	{N}^{vc}_{0} = {N}_{0} $. Therefore, probability of infection remains unchanged.  

The general form of the probability of infection under vaccinated and unvaccinated situations can be expressed as 

\begin{equation}
	P = 1- e^{(-\alpha \frac{N}{N_0})},
\end{equation}

where $ \alpha=1 $ for no vaccination cases and $ \alpha = 1- \eta_{vc} $ for vaccinated cases. 

The net probability of safety decreases for the variants strains due to higher transmissibility $ \tau $ compared to the original strain. With $ \hat{P}^{vr}_{net} $ representing the probability of no infection, its relationship with $ \hat{P}_{net} $ can be written as  

\begin{equation}
	\hat{P}^{vr}_{net} = \frac{\hat{P}_{net}}{\tau}
\end{equation}

Along the same lines of derivation of $ \alpha $ for the vaccination case, it can be shown that $ \alpha = \tau $.

\end{document}